\documentclass[a4paper,12pt]{article}
\usepackage{cite}
\usepackage{epsfig}
\usepackage{amssymb,amsmath}
\usepackage{graphicx}
\usepackage{float}
\usepackage{comment}
\usepackage{epstopdf}

\usepackage{caption3} 
\DeclareCaptionOption{parskip}[]{} 
\usepackage[small]{caption}

\pdfoutput=1

\newlength{\dinwidth}
\newlength{\dinmargin}
\newcommand{\cO}{\mathcal{O}}
\newcommand{\cR}{\mathcal{R}}
\newcommand{\eqn}[1]{(\ref{#1})}
\newcommand{\bel}[1]{\be\label{#1}}
\newcommand{\e}{\mathrm{e}}

\newcommand{\ta}[0]{\tilde \alpha}

\def\be{\begin{equation}}
\def\ee{\end{equation}}
\def\beqn{\begin{eqnarray}}
\def\eeqn{\end{eqnarray}}
\def\no{\nonumber}
\def\ba{\begin{array}{c}}
\def\bat{\begin{array}{cc}}
\def\ea{\end{array}}
\def\bi{\begin{itemize}}
\def\ei{\end{itemize}}

\def\cL{{\cal L}}

\def\cM{{\cal M}}

\def\cO{{\cal O}}

\def\be{\begin{equation}}
\def\ee{\end{equation}}

\begin{document}

\title{
\begin{flushright}\vbox{\normalsize \mbox{}\vskip -6cm
FTUV/14-0526 \\[-3pt] IFIC/14-36
}
\end{flushright}\vskip 45pt
{\bf Low-mass fermiophobic charged Higgs phenomenology in two-Higgs-doublet models}}
\bigskip

\author{Victor Ilisie and Antonio Pich \\[15pt]
{\small IFIC, Universitat de Val\`encia -- CSIC, Apt. Correus 22085, E-46071 Val\`encia, Spain}}

\date{}  
\maketitle
\bigskip \bigskip

\begin{abstract}

After the recent discovery of a Higgs-like boson, the possibility of an enlarged scalar sector arises as a natural question. Experimental searches for charged scalars have been already performed with negative results. We analyze the phenomenology associated with a fermiophobic charged Higgs (it does not couple to fermions at tree level),
in two-Higgs-doublet models. All present experimental bounds are evaded trivially in this case, and one needs to consider other decay and production channels. We study the associated production of a charged Higgs with either a $W$ or a neutral scalar boson, and the relevant decays for a light fermiophobic charged Higgs. The interesting features of this scenario should result encouraging for the LHC collaborations to perform searches for such a particle.

\end{abstract}

\newpage

\section{Introduction}
\label{sec:introd}

The recent discovery of a boson with mass around 125 GeV by the ATLAS \cite{Aad:2012tfa,Aad:2013wqa,AtlasConf1,AtlasConf2}, CMS \cite{Chatrchyan:2012ufa,Chatrchyan:2013lba,CMSpas}, D\O~and CDF \cite{Aaltonen:2012qt,Aaltonen:2013kxa} collaborations is the first direct hint of the electroweak symmetry-breaking mechanism. The experimental data
confirm that it is a Higgs-like scalar with couplings compatible with the Standard Model (SM) predictions. However,
this new particle could belong to an enlarged scalar sector.

In order to give mass to fermions and gauge bosons while preserving gauge invariance, the SM assumes the presence of one SU(2) electroweak scalar doublet with a non-zero vacuum expectation value. However, no fundamental principle or symmetry forbids the presence of additional scalar doublets.
The simplest extension of the SM is the two-Higgs-doublet model (2HDM)~\cite{Gunion:1989we,Branco:2011iw}, which leads to
a richer scalar sector and very interesting phenomenological implications \cite{ilisie3,ilisie2,ilisie1,Pich:2009sp,JungTuzon,PichTuzon1,Jung:2012vu,PichDstar,PichEDP,PichBll,LHCP2013,
SDG,Barroso:2013zxa,
Grinstein:2013npa,Eberhardt:2013uba,Chen:2013rba,Craig:2013hca,Coleppa:2013dya,Shu:2013uua,
Chiang:2013ixa,Krawczyk:2013jta,SK:13,IDM1,IDM2,Belanger:2013xza,Enberg:2013ara,Dorsch:2013wja,
Dorsch:2014qja,Bhattacharyya:2014nja,Altmannshofer:2012ar,Chang:2013ona,Cheung:2013rva,Enberg:2013jba}.
Generic multi-Higgs doublet models give rise to unwanted flavour-changing neutral current (FCNC) interactions, which are found to be very suppressed experimentally. The FCNCs can be eliminated at tree level by requiring the alignment in flavour space of the Yukawa matrices \cite{Pich:2009sp}. The so-called aligned two-Higgs-doublet model (A2HDM) contains as particular cases the different versions of 2HDMs with discrete $\mathcal{Z}_2$ symmetries while at the same time introduces new sources of CP violation beyond the CKM phase.

The main feature of the 2HDM is the presence of three neutral and one charged Higgs bosons.
Finding extra neutral or charged scalar bosons would be a clear signal of an extended scalar sector.
The ATLAS \cite{Aad:2012tj,Aad:2013hla} and CMS collaborations \cite{Chatrchyan:2012vca} have performed direct searches for a charged Higgs particle. However, since no excess has been found over the SM background, this only allows us to further constrain the parameter space of the various types of 2HDMs; recent analyses within the A2HDM have been performed in \cite{ilisie3,ilisie2,ilisie1}. In their searches, both collaborations assume
that the charged Higgs is produced in a top-quark decay ($t\to H^+ b$) and that it decays dominantly into fermions; {\it i.e.}, $H^+ \to q_u \bar{q}_d, \; l^+ \nu_l$. However, all experimental bounds would be trivially evaded for a fermiophobic charged Higgs, {\it i.e.}, a charged scalar which does not couple to fermions at tree level.
In order to probe such scenario, other production channels and decay rates would have to be considered.
Although such analyses have not been yet performed by the LHC collaborations, they become more compelling as the experimental bounds on a non-fermiophobic charged Higgs are getting stronger, at least in the low mass region.
The fermiophobic scenario is a simplified model that, if it turns out to be the one preferred by Nature, would allow us to measure (or at least estimate) for the first time the parameters of the scalar potential. This is usually a rather difficult task in more generic 2HDM settings.
It is also worth mentioning that a fermiophobic charged Higgs is present in the {\it inert\/} 2HDM~\cite{Ma:2008uza,Ma:2006km}, where one of the neutral scalars is a nice candidate for dark matter
\cite{Krawczyk:2013jta,
SK:13,IDM1,IDM2,Enberg:2013ara,CMR:07,Barbieri:2006dq,LopezHonorez:2006gr,ALT:09,LHY:10,LHY:11,
DS:09,Arhrib:2013ela,Ginzburg:2010wa}. The discovery of a fermiophobic $H^\pm$ particle could be interpreted in this case as an indirect signal of the presence of dark matter.

In this work, we shall focus our analysis on the search of a light fermiophobic charged Higgs $H^\pm$, with mass in the range $M_{H^\pm}\in [M_W,M_W+M_Z]$ so that only a few relevant decay modes are kinematically open.
We will study the two most important production channels for a fermiophobic $H^\pm$: associated production with either a $W^\mp$ boson or a neutral scalar.
Due to their similarity with the SM Higgs production channels, one expects them to be experimentally accessible at LHC energies. Next-to-leading order (NLO) QCD corrections will be included for both cross sections, and the bounds on the various parameters of the model from the current LHC data \cite{ilisie3} will also be taken into account.
The main features of the A2HDM are briefly presented in section 2. Section 3 discusses the calculation of the various decay rates and production modes. Finally, in section 4 we perform a phenomenological analysis, assuming different scenarios for the scalar spectrum, and conclude in section 5 with a summary of our results. Some technical details are given in four appendices.


\section{The Aligned Two-Higgs-Doublet Model}
\label{sec:A2HDM}

The 2HDM extends the SM with a second scalar doublet of hypercharge $Y=\frac{1}{2}$.
The neutral components of the two scalar doublets acquire vacuum expectation values that are in general complex,
$\langle 0|\phi_a^{(0)}(x)|0\rangle =\frac{1}{\sqrt{2}}\, v_a\, \e^{i\theta_a}$ ($a=1,2$), although
only the relative phase $\theta \equiv \theta_2 - \theta_1$ is observable.
It is convenient to perform a global SU(2) transformation in the scalar space $(\phi_1,\phi_2)$,
characterized by the angle $\beta = \arctan{(v_2/v_1)}$,
and work in the so-called Higgs basis
$(\Phi_1,\Phi_2)$, where only one doublet acquires a vacuum expectation value:
\begin{equation}  \label{Higgsbasis}
\Phi_1=\left[ \begin{array}{c} G^+ \\ \frac{1}{\sqrt{2}}\, (v+S_1+iG^0) \end{array} \right] \; ,
\qquad\qquad\qquad
\Phi_2 = \left[ \begin{array}{c} H^+ \\ \frac{1}{\sqrt{2}}\, (S_2+iS_3)   \end{array}\right] \; ,
\end{equation}
where $G^\pm$ and $G^0$ denote the Goldstone fields.
Thus, $\Phi_1$ plays the role of the SM scalar doublet with
$v\equiv \sqrt{v_1^2+v_2^2}\simeq (\sqrt{2}\, G_F)^{-1/2} = 246~\mathrm{GeV}$.

The physical scalar spectrum contains five degrees of freedom: the two charged fields $H^\pm(x)$
and three neutral scalars $\varphi_i^0(x)=\{h(x),H(x),A(x)\}$, which are related with the $S_i$ fields
through an orthogonal transformation $\varphi^0_i(x)=\mathcal{R}_{ij} S_j(x)$.
The form of the $\mathcal{R}$ matrix is fixed by the scalar potential \cite{ilisie1}, which determines the neutral scalar mass matrix
and the corresponding mass eigenstates. A detailed discussion is given in appendix \ref{app:potential}. In general, the CP-odd component $S_3$ mixes with the CP-even fields
$S_{1,2}$ and the resulting mass eigenstates do not have a definite CP quantum number.
If the scalar potential is CP symmetric this admixture disappears; in this particular case, $A(x) = S_3(x)$
and
\bel{eq:CPC_mixing}
\left(\ba h\\ H\ea\right)\; = \;
\left[\bat \cos{\tilde\alpha} & \sin{\tilde\alpha} \\ -\sin{\tilde\alpha} & \cos{\tilde\alpha}\ea\right]\;
\left(\ba S_1\\ S_2\ea\right) \, .
\ee
Performing a phase redefinition of the neutral CP-even fields, we can fix the sign of $\sin{\ta}$.  In this work we adopt the conventions\ $M_h \le M_H$\ and\
$ 0 \leq \ta \leq \pi$, so that $\sin{\ta}$ is positive.

The most generic Yukawa Lagrangian with the SM fermionic content gives rise to FCNCs because the fermionic couplings of the two scalar doublets cannot be simultaneously diagonalized in flavour space. The non-diagonal neutral couplings can be eliminated by requiring the alignment in flavour space of the Yukawa matrices~\cite{Pich:2009sp}; {\it i.e.}, the two Yukawa matrices coupling to a given type of right-handed fermions are assumed to be proportional to each other and can, therefore, be diagonalized simultaneously. The three proportionality parameters $\varsigma_f$~($f=u,d,l$) are arbitrary complex numbers and introduce new sources of CP violation.
In terms of the fermion mass-eigenstate fields, the Yukawa interactions of the A2HDM read~\cite{Pich:2009sp}
\beqn\label{lagrangian}
 \mathcal L_Y & = &  - \frac{\sqrt{2}}{v}\; H^+ \left\{ \bar{u} \left[ \varsigma_d\, V M_d \mathcal P_R - \varsigma_u\, M_u^\dagger V \mathcal P_L \right]  d\, + \, \varsigma_l\, \bar{\nu} M_l \mathcal P_R l \right\}
\nonumber \\
& & -\,\frac{1}{v}\; \sum_{\varphi^0_i, f}\, y^{\varphi^0_i}_f\, \varphi^0_i  \; \left[\bar{f}\,  M_f \mathcal P_R  f\right]
\;  + \;\mathrm{h.c.} \, ,
\eeqn
where $\mathcal P_{R,L}\equiv \frac{1\pm \gamma_5}{2}$ are the right-handed and left-handed chirality projectors,
$M_f$ the diagonal fermion mass matrices
and the  couplings of the neutral scalar fields are given by:
\begin{equation}    \label{yukascal}
y_{d,l}^{\varphi^0_i} = \cR_{i1} + (\cR_{i2} + i\,\cR_{i3})\,\varsigma_{d,l}  \, ,
\qquad\qquad
y_u^{\varphi^0_i} = \cR_{i1} + (\cR_{i2} -i\,\cR_{i3}) \,\varsigma_{u}^* \, .
\end{equation}
As in the SM, all scalar-fermion couplings are proportional to the corresponding fermion masses, and
the only source of flavour-changing interactions is the Cabibbo-Kobayashi-Maskawa~(CKM) quark mixing matrix $V$~\cite{Cabibbo:1963yz,Kobayashi:1973fv}.
The usual models with natural flavour conservation, based on discrete ${\cal Z}_2$ symmetries, are recovered for particular (real) values of the couplings $\varsigma_f$~\cite{Pich:2009sp}.

The full set of interactions among the gauge and scalar bosons is given in \cite{ilisie1}. The coupling of a single neutral scalar with a pair of gauge bosons takes the form ($V=W,Z$)
\begin{align}
g_{\varphi_i^0 VV} = \mathcal{R}_{i1} \; g^{\text{SM}}_{hVV}\, ,
\end{align}
which implies $g_{hVV}^2 + g_{HVV}^2 + g_{AVV}^2 = (g_{hVV}^\text{SM})^2$. Thus, the strength of the SM Higgs interaction is shared by the three 2HDM neutral bosons. In the CP-conserving limit, the CP-odd field decouples while the strength of the $h$ and $H$ interactions is governed by the corresponding $\cos\tilde\alpha$ and $\sin\tilde\alpha$ factors.

In the following analysis we are also going to need the coupling of a neutral scalar with a pair of charged Higgses. We have parametrized the corresponding interaction as:
\be
\cL_{\varphi^0 H^+H^-}\; =\; - v \;\sum_{\varphi^0_i}\, \lambda_{\varphi^0_i H^+H^-}\;\, \varphi^0_i\, H^+H^-\, .
\label{hHPHM}
\ee
Explicit expressions for the reduced cubic couplings $\lambda_{\varphi^0_i H^+ H^-}$, in terms of the generic Higgs potential parameters, can be found in \cite{ilisie1}.

The phenomenological constraints on the A2HDM parameters have been studied in detail in Refs.
\cite{ilisie3,ilisie2,ilisie1,Pich:2009sp,JungTuzon,PichTuzon1,Jung:2012vu,PichDstar,PichBll,PichEDP}. For a light $H^\pm$, loop-induced processes dominated by top contributions ($\varepsilon_K$, $Z\to b\bar b$, $B^0$--$\bar B^0$ mixing) impose a tight (95\% CL) upper bound on the up-type alignment parameter: $|\varsigma_u| < 0.77 \; (1.7)$, for $M_{H^\pm} = 80$ (500) GeV. Owing to the much smaller fermion masses, the constraints on the down-type (and lepton) parameter are very weak; one imposes instead $|\varsigma_d| \le 50$ to guarantee a perturbative Yukawa coupling. In the popular type-II 2HDM
($\varsigma_u = -1/\varsigma_d = -1/\varsigma_l =\cot\beta$), the decay $\bar B\to X_s\gamma$ excludes charged Higgs masses below 380 GeV \cite{Hermann:2012fc} at 95\% CL, because the SM and charged-Higgs contributions interfere constructively. This is no longer true in the more general A2HDM framework, where one only gets a combined correlated constraint on $M_{H^\pm}$, $\varsigma_u$ and $\varsigma_d$, which allows much lighter values of the charged-scalar mass in a restricted region of the parameter space $\varsigma_u$--$\varsigma_d$ \cite{JungTuzon,PichTuzon1,Jung:2012vu}.

The symmetries of the A2HDM protect in a very efficient way the flavour-blind phases of the alignment parameters from undesirable phenomenological consequences. The experimental upper bounds on fermion electric dipole moments provide the
strongest constraints on $\mathrm{Im}(\varsigma_f)$, but $\cO(1)$ contributions remain allowed at present \cite{PichEDP}.
For simplicity, in section \ref{sec:phenom}, we will restrict our analysis to the CP-conserving limit and, therefore, will consider real alignment parameters.
The LHC data require the gauge coupling of the 125~GeV boson to have a magnitude close to the SM one. Assuming that it corresponds to the lightest CP-even scalar $h$ of the CP-conserving A2HDM, the measured Higgs signal strengths imply $|\cos{\tilde\alpha}| > 0.90\; (0.80)$ at 68\% (90\%) CL \cite{ilisie3,ilisie2,ilisie1}.
Direct searches for a heavier neutral scalar ($H$) provide upper bounds on $|\sin{\tilde\alpha}|$ as a function of $M_H$, which at present result in a weaker
constraint on the mixing angle \cite{ilisie3}.

In the following we will explore the intriguing possibility that the charged scalar could be fermiophobic, {\it i.e.}, that its tree-level couplings to fermions vanish ($\varsigma_{u,d,l}=0$). All current experimental bounds are then trivially avoided,
in particular the flavour constraints \cite{JungTuzon}. The Yukawa couplings of the h(125) boson scale in this case, with respect to the SM ones, with the same factor as the gauge couplings: $y^h_f = {\cal R}_{11} = \cos{\tilde\alpha}$. The global fit to the Higgs signal strengths results in the slightly improved bound $|\cos{\tilde\alpha}| > 0.86$ at 90\% CL \cite{ilisie3}.

In the fermiophobic (and CP-conserving) limit, the CP-odd scalar $A$ has also vanishing Yukawa couplings. Therefore, it only couples via multi-Higgs interactions with an even number of $A$ bosons, or through its gauge couplings
($A W^\pm H^\mp$, $A Z h$, $A Z H$, $A^2Z^2$, $A^2W^+W^-$, $AH^\pm W^\mp \gamma$, $AH^\pm W^\mp Z$).
Thus, a light $A$ boson might be very long-lived. While this could have cosmological implications, it is not in conflict
with the relic-density constraints
\cite{Krawczyk:2013jta,SK:13,CMR:07,Barbieri:2006dq,LopezHonorez:2006gr,ALT:09,LHY:10,LHY:11,DS:09,Arhrib:2013ela,Ginzburg:2010wa}.

A more specific version of the fermiophobic scenario is provided by the {\it inert\/} 2HDM~\cite{Ma:2008uza,Ma:2006km}, which assumes a discrete ${\cal Z}_2$ symmetry in the Higgs basis such that all SM fields and $\Phi_1$ are even ($\Phi_1\to \Phi_1$) under this symmetry while the second (inert) scalar doublet is odd ($\Phi_2\to -\Phi_2$). In this restricted case, there is no mixing between the CP-even neutral scalars $h$ and $H$; {\it i.e.}, $\cos{\tilde\alpha} = 1$. The spectrum of the {\it inert\/} 2HDM is described in appendix~\ref{subsec:inert}.

\section{Decay and Production modes}
\label{sec:calc}

We are going to analyse the possibility of having a fermiophobic charged Higgs with a mass in the restricted interval $M_{H^\pm} \in [M_W, \!\ M_W+M_Z]$. In this region, the only relevant decay rates are $H^+\to W^+\gamma$ and $H^+\to W^+\varphi_i^0$.
We are mainly interested in the one-loop suppressed decay
$H^+\to W^+\gamma$, the only two-body kinematically allowed decay mode, but we need to account also for the tree-level decay into a $W^+$ boson and a neutral scalar, which cannot be both on-shell simultaneously for the whole considered kinematical region. Thus,
we shall consider three-body decays like $H^+\to W^+ f\bar{f}$ mediated by the neutral scalars $\varphi_i^0$
and $H^+\to \varphi_i^0 f_u\bar{f}_d$ mediated by a virtual $W^+$,
where $f_u\bar{f}_d$ stands for quark pairs $q_u\bar{q}_d$, or lepton-neutrino pairs $l^+\nu_l$. The loop-induced decay $H^+\to f_u\bar{f}_d$ has a strong Yukawa suppression $m_f^2/v^2$ and, therefore, it is irrelevant for this discussion.
When surpassing the $M_W+M_Z$ threshold, the one-loop decay $H^+\to W^+Z$ would enter the game and we would also be close to the top-quark production threshold. The analysis of these two extra decay modes lays beyond the goal of this paper.

\subsection{$\mathbf{H^+\to W^+\gamma}$}

The first process that we are going to analyse is $H^+(k+q)\to W^+(k)\,\gamma(q)$.
Owing to the conservation of the electromagnetic current, the
decay amplitude must adopt the form:
\begin{align}
\cM\; =\;\Gamma^{\mu \nu}\,\varepsilon_\mu^*(q)\,\varepsilon_\nu^*(k)\, ,
\qquad\qquad
\Gamma^{\mu \nu}\; =\;  \left( g^{\mu\nu} k \cdot q - k^\mu q^\nu\right) \,  S \, +  \, i \,  \epsilon^{\mu\nu\alpha\beta} \,  k_\alpha \, q_\beta \;  \tilde{S}\, ,
\label{transverse}
\end{align}
where $S$ and $\tilde S$ are scalar form factors.
To obtain this expression,
we have considered the most general Lorentz structure for the effective $\Gamma^{\mu\nu}$ vertex, and have imposed the electromagnetic current conservation condition $q_\mu\, \Gamma^{\mu\nu} = 0$.
All terms proportional to $q^\mu$ and $k^\nu$ have been also eliminated, as they cancel when contracted with the polarization vectors of the photon and the $W$ boson. Note that, accidentally, the Ward-like identity $k_\nu \,\Gamma^{\mu\nu}=0$ also holds for (\ref{transverse}).

\begin{figure}[t]
\centering
\includegraphics[scale=0.44]{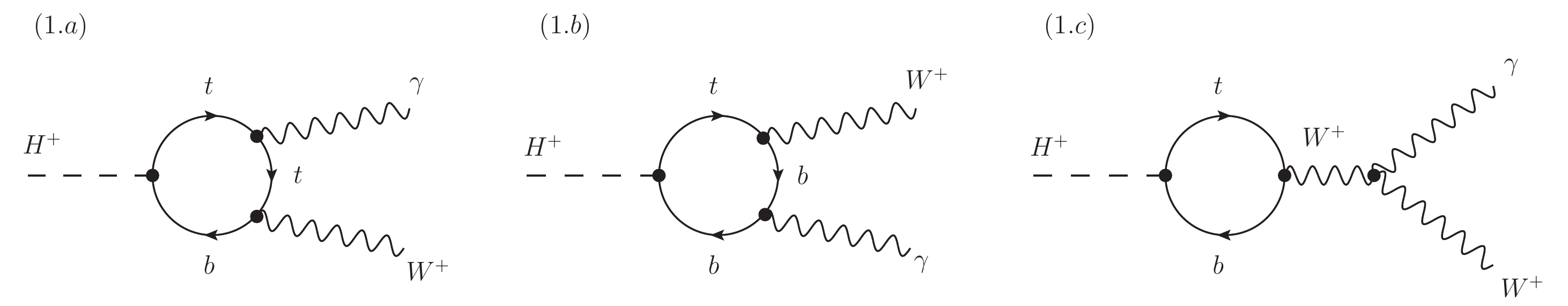} \\
\includegraphics[scale=0.44]{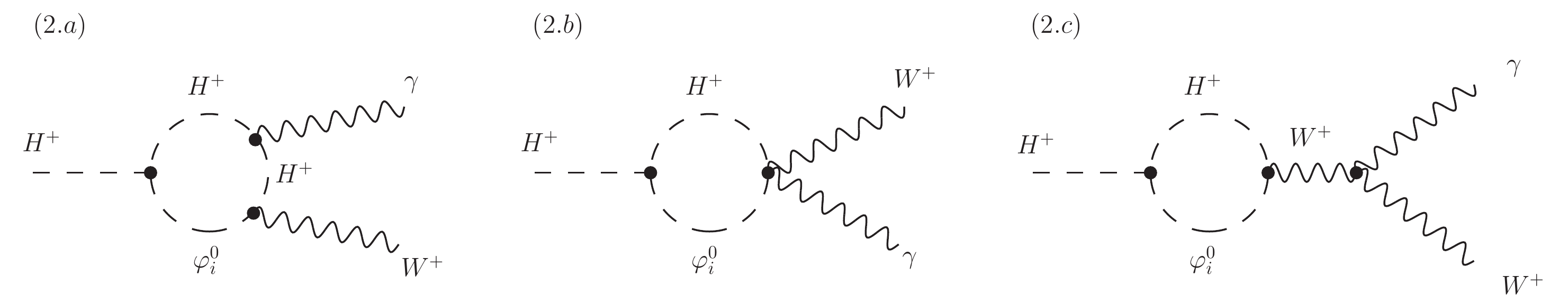} \\
\includegraphics[scale=0.44]{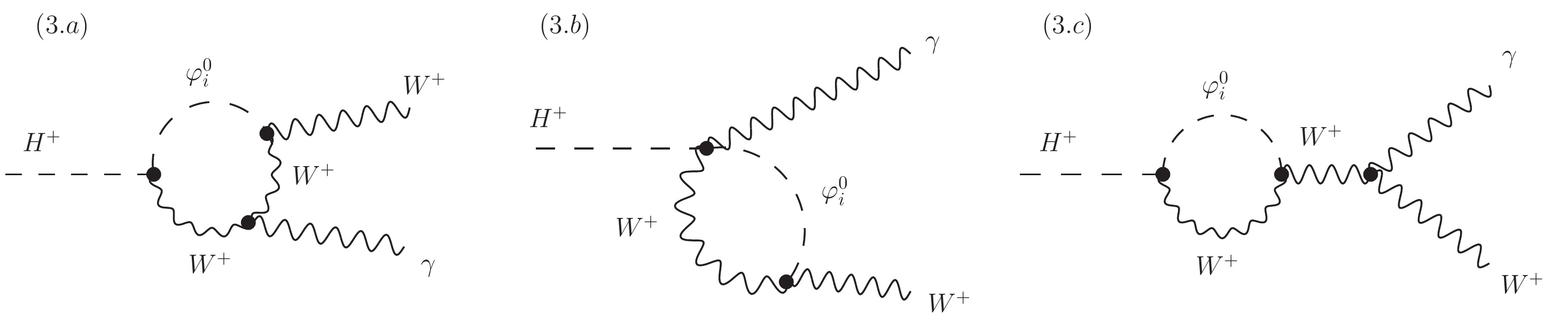}
\caption{\it One-loop diagrams contributing to $H^+\to W^+\gamma$ in the unitary gauge.}
\label{oneLoop}
\end{figure}

In the unitary gauge, the decay proceeds at one loop through
the three sets of diagrams shown in Fig.~\ref{oneLoop}: fermionic loops (set 1), scalar loops (set 2) and loops with both gauge and scalar bosons (set 3).
Each set is transverse by itself, {\it i.e.}, of the form given in (\ref{transverse}). We can then decompose the result into the three separate contributions:
$S = S_{(1)} + S_{(2)} + S_{(3)}$ and $\tilde{S}=\tilde{S}_{(1)}$ (the only contribution to the structure $\epsilon^{\mu\nu\alpha\beta} \!\ k_\alpha \!\ q_\beta$ comes from the fermionic loops).
One can further simplify the calculation of $S_{(j)}$
by only considering the terms of the transverse set $j$ that contribute to the structure $k^\mu q^\nu$.
In order to calculate these contributions, one only needs to compute diagrams 1.a and 1.b for the first set, 2.a for the second set and 3.a for the third one.

\noindent We obtain the following expressions for the form factors:
\begin{align}
S_{(1)}\; = &\;\; \frac{\alpha \!\ N_C  \!\ |V_{tb}|^2}{2\pi \!\ v \!\  s_{_\text{W}}}\;  \int_0^1 dx \int_0^1 dy  \;\, \left[ Q_t\, x + Q_b\, (1-x)\right]
\notag \\ &\;\; \times \;\; \frac{-\varsigma_u m_t^2 \, x\, (2xy -2y + 1) \, + \, \varsigma_d m_b^2 \, (1-x)(1-2xy)}{M_W^2\, x\, (x-1) + m_b^2\, (1-x)+ m_t^2\, x + (M_W^2-M_{H^\pm}^2)\, xy\, (1-x)} \; ,
\\[2.2ex]
S_{(2)} \; = &\;\; \frac{\alpha\, v}{2\pi \, s_{_\text{W}}} \; \sum_i \; \lambda_{\varphi_i^0 H^+ H^-}  \!\  \big(\mathcal{R}_{i2}- i\mathcal{R}_{i3} \big)  \; \int_0^1 dx \int_0^1 dy
\notag \\ & \;\; \times  \;\;  \frac{x^2 y\, (1-x)}{M_W^2\, x\, (x-1) + M_{\varphi_i^0}^2\, (1-x)+ M_{H^\pm}^2\, x + (M_W^2-M_{H^\pm}^2)\, xy\, (1-x)} \; ,
\\[2.2ex]
S_{(3)}\; = &\;\; \frac{\alpha}{2\pi v  \, s_{_\text{W}}}\; \sum_i \;  \mathcal{R}_{i1} \big(\mathcal{R}_{i2}- i\mathcal{R}_{i3} \big)  \; \int_0^1 dx \int_0^1 dy  \;\,  x^2
\notag \\ & \;\;\times \;\;
\frac{ 2M_W^2 \, + \, \big(M_{H^\pm}^2 + M_W^2 - M_{\varphi_i^0}^2\big) \, y\, (x-1)}{M_W^2\, x^2 + M_{\varphi_i^0}^2\, (1-x)+ (M_W^2-M_{H^\pm}^2)\, xy\, (1-x)} \; ,
\\[2.2ex]
\tilde{S} \; = &\;\; \frac{\alpha\, N_C  \, |V_{tb}|^2}{2\pi \, v \,  s_{_\text{W}}}  \int_0^1 dx \int_0^1 dy  \;\,
\left[ Q_t\, x + Q_b\, (1-x)\right]
\notag \\ & \;\; \times \;\;
\frac{\varsigma_u m_t^2 \, x\, +\,\varsigma_d m_b^2\, (1-x)}{M_W^2\, x\, (x-1) + m_b^2\, (1-x)+ m_t^2\, x + (M_W^2-M_{H^\pm}^2)\, xy\, (1-x)} \; ,
\end{align}
with $s_{_\text{W}}\equiv\sin{\theta_{_{\mathrm{W}}}}$.
The calculation of $S_{(3)}$ has been also performed in the Feynman ($\xi=1$) gauge,
where additional diagrams with Goldstone bosons are present, verifying that these expressions are gauge independent. Our results are
in agreement with the recent calculation of the $H^+W^-\gamma$ effective vertex in Ref.~\cite{EffVertex}. This calculation was also done many years ago by several groups \cite{OldHWgamma1,OldHWgamma2,OldHWgamma3,OldHWgamma4} using a somewhat different notation.

The $H^+\to W^+\gamma$ decay width is easily found to be:
\be
\Gamma(H^+\to W^+\gamma)\; =\; \frac{M_{H^\pm}^3}{32 \pi} \; \left(1-\frac{M_W^2}{M_{H^\pm}^2}\right)^3 \, \left(\, |S|^2+|\tilde{S}|^2 \,\right)\, .
\ee
This one-loop decay rate is in general much smaller than the tree-level decay rates of a charged Higgs into fermions. However, it becomes relevant if the charged Higgs is fermiophobic ($\varsigma_f\to 0$). In this case, the first set of diagrams (which has only been presented for completeness) does not contribute.

\subsection{$\mathbf{H^+\to W^+\varphi_i^0}$}

\begin{figure}[t]
\centering
\includegraphics[scale=0.6]{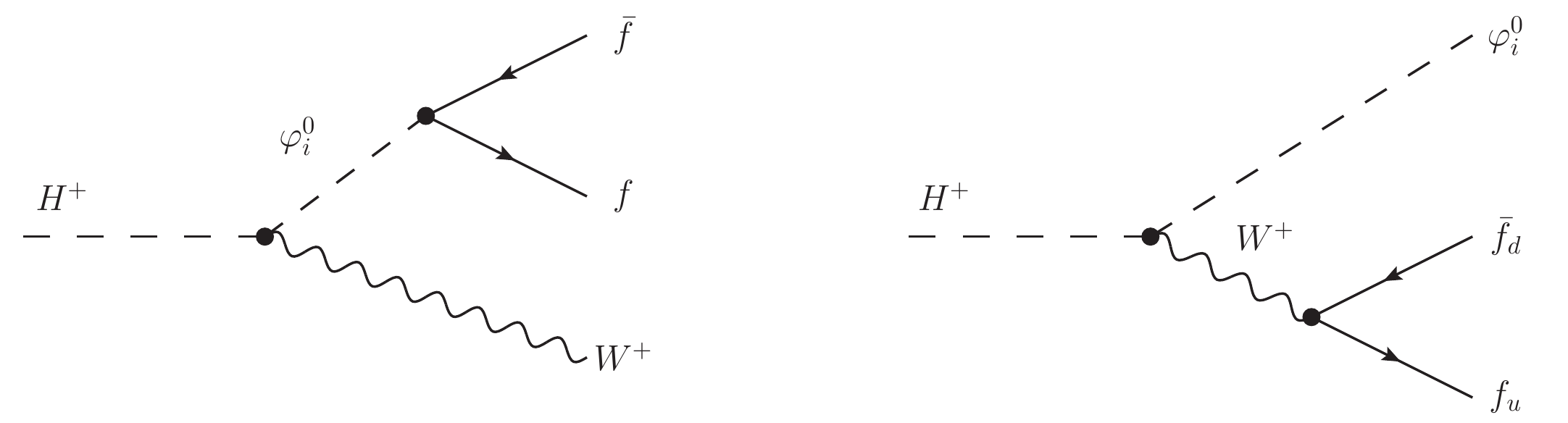}
\caption{\it $H^+\to W^+ f\bar{f}$ process mediated by the virtual neutral scalars $\varphi_i^0$ (left) and $H^+\to \varphi_i^0 f_u\bar{f}_d$ mediated by a virtual $W^+$ (right).}
\label{threebody}
\end{figure}

The $H^+$ decay rate to on-shell $W^+$ and $\varphi_i^0$ bosons is given by
\be
\Gamma(H^+ \to W^+\varphi_i^0) \; = \;
\frac{\alpha}{16\, s_{_\text{W}}^2\,M_{H^\pm}^3 \, M_W^2} \;
\left(\mathcal{R}_{i2}^2 +  \mathcal{R}_{i3}^2\right) \; \lambda^{3/2}( M_{\varphi_i^0}^2, M_{H^\pm}^2,M_W^2) \, ,
\ee
with the usual definition of the lambda function $\lambda(x,y,z)\equiv x^2+y^2+z^2-2xy-2xz-2yz$.

The corresponding three-body decay rate to $W^+f\bar{f}$, with off-shell neutral scalars (Fig.~\ref{threebody}, left), takes the form:
\begin{align}
 \Gamma(H^+ \to W^+ f \bar{f} )\; & = \;
 \frac{\alpha^2\, N_C^f\, m_f^2}{128\,\pi\, s_{_{\text{W}}}^4 \, M_{H^\pm}^3 \, M_W^4}\;
 \int_{4 m_f^2}^{(M_{H^\pm}-M_W)^2} ds_{23} \;\;
 \lambda^{3/2}(M_{H^\pm}^2,M_W^2,s_{23})
 \notag \\[1.5ex]
 & \qquad\qquad \times \; \left( 1-\frac{4 m_f^2}{s_{23}} \right)^{1/2} \; \sum_{i,j}  \;
 \big(\mathcal{R}_{i2}-i\mathcal{R}_{i3}\big)
\big(\mathcal{R}_{j2}+i\mathcal{R}_{j3}\big) \; \mathcal{M}_{ij} \; ,
\end{align}
where $N_C^f$ stands for the number of colours of the fermion $f$, 3 for quarks and 1 for leptons, $s_{23}$ is the square of the fermion-antifermion invariant mass and
\begin{align}
 \mathcal{M}_{ij} \; \equiv \; \frac{(s_{23}-2m_f^2) \; \text{Re}\big(y_f^{\varphi_i^0}  y_f^{\varphi_j^0 *} \big) - 2m_f^2 \; \text{Re}\big(y_f^{\varphi_i^0}  y_f^{\varphi_j^0 }\big)}{(s_{23}-M_{\varphi_i^0}^2)(s_{23}-M_{\varphi_j^0}^2)}
    \; .
\end{align}
Obviously, the $b$-quark contribution will dominate because of the global factor $m_f^2$. Therefore, we will neglect the other fermionic final states.

For the decay $H^+\to \varphi_i^0 f_u\bar{f}_d$, with an of-shell $W^+$ (Fig.~\ref{threebody}, right), we are going to consider all possible final states, quarks and leptons. We exclude the top quark, since this process is well below its production threshold. Neglecting the final fermion masses, the sum over all kinematically-allowed decay modes amounts to a global factor
\be
\Omega\; = \; \left( 3 + N_C \sum_{u_i = u,c} \,\sum_{d_j = d,s,b} |V_{u_i d_j}|^2 \right) \; = \; 9 \, ,
\ee
where the unitarity of the CKM matrix has been used. The total decay width can be expressed as an integral over the fermion-antifermion invariant-mass squared:
\begin{align}
\Gamma\Bigl(H^+ \to \varphi_i^0 \sum_{f_u,f_d} f_u\bar{f}_d \Bigr) \; = \; &
 \frac{\Omega}{9} \;
 \frac{3 \,\alpha^2\, (\mathcal{R}_{i2}^2 + \mathcal{R}_{i3}^2 )}{64\,\pi\, s_{_{\text{W}}}^4 \, M_{H^\pm}^3}\;
 \int_{0}^{(M_{H^\pm}-M_{\varphi_i^0})^2} \; ds_{23}  \;  \frac{\lambda^{3/2}(M_{H^\pm}^2,M_{\varphi_i^0}^2,s_{23})}{(s_{23}-M_W^2)^2} .
\end{align}
%


\subsection{Charged-Higgs Production}

\begin{figure}[t]
\centering
\includegraphics[scale=0.42]{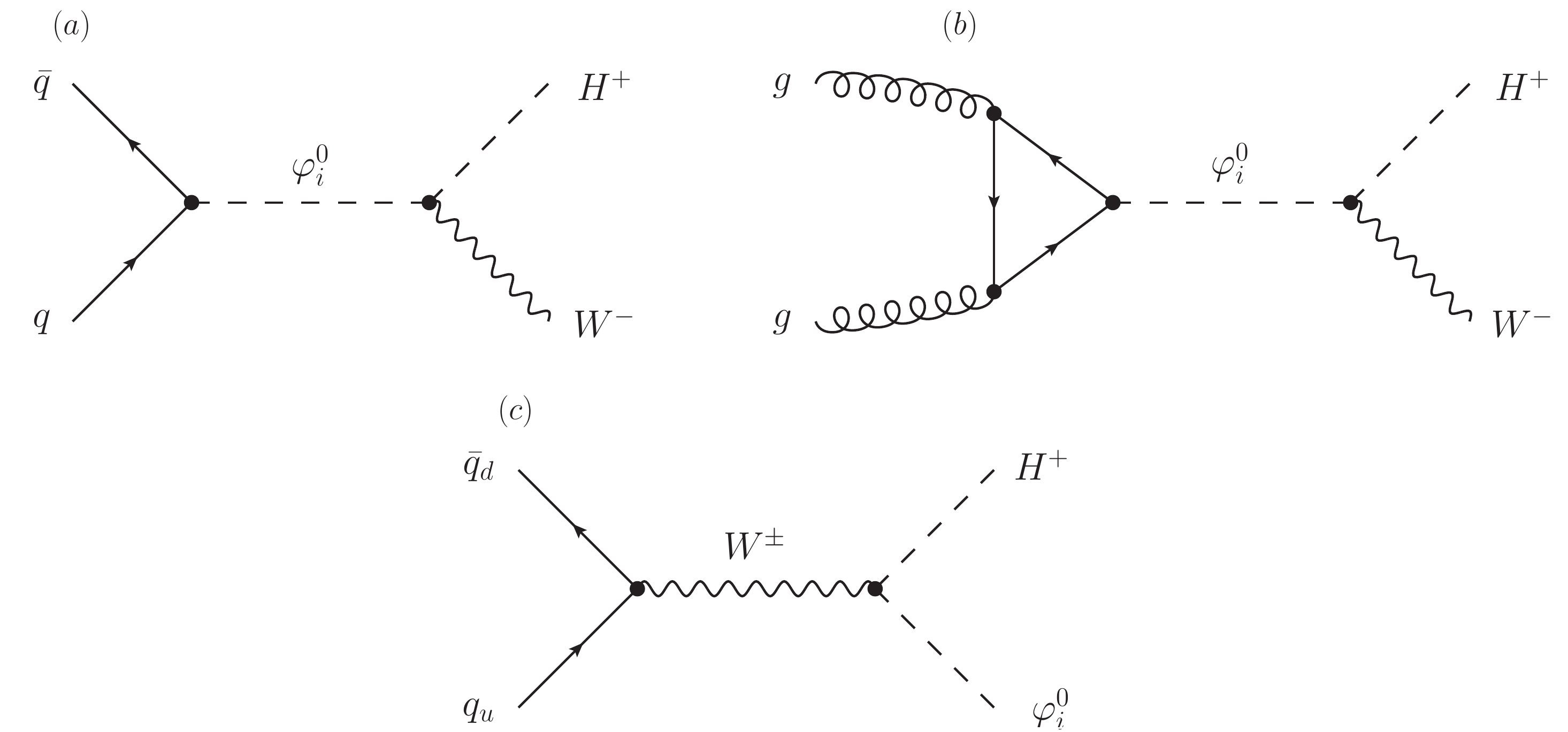}
\caption{\it LO contributions to the charged-Higgs associated production with a $W$ boson (diagrams a, b) or a neutral scalar (diagram c), in the fermiophobic scenario.}
\label{productionCH}
\end{figure}

In order to see if the fermiophobic scenario can be experimentally probed, one needs an estimation of the production cross sections for different channels. Here we
will consider two possibilities, the associated production with a neutral scalar and the associated production with a $W$ boson (Fig.~\ref{productionCH}).
The $q_u\bar{q}_d\to H^+\varphi_i^0$ production process is by far the most interesting channel, as it requires the least number of new parameters.
For initial-state massless quarks, the leading-order (LO) partonic cross section reads
\be
\hat{\sigma}(q_u\bar{q}_d\to H^+\varphi_i^0) \; = \; \frac{g^4 \; |V_{ud}|^2}{768 \; \pi \; N_c \; \hat{s}^2} \;
\frac{(\mathcal{R}_{i2}^2+\mathcal{R}_{i3}^2)}{(\hat s - M_W^2)^2} \; \lambda^{3/2}(\hat{s},M_{H^\pm}^2,M_{\varphi_i^0}^2) \; ,
\label{drell-yan}
\ee
where $\hat s$ is the partonic invariant-mass squared.
The NLO QCD corrections are available and can be expressed in a very simple form, as shown in appendix~\ref{QCDcorrections}.

The associated production with a $W$ boson can proceed through either $q\bar{q}$ or
$gg$ fusion. The partonic LO cross section for the $q\bar{q}$ fusion process, is given by
\begin{align}
\hat{\sigma}(q\bar{q}\to H^+ W^-)\; &=\; \frac{g^2}{128 \; \pi \, M_W^2 \, {\hat s}^2} \; \frac{m_q^2}{v^2} \; \frac{1}{N_c} \;
 \lambda^{3/2}(\hat s,M_{H^\pm}^2,M_W^2) \;\left(1 -\frac{4 m_q^2}{\hat s}\right)^{-1/2}
\notag  \\
& \qquad\qquad\qquad \times \; \sum_{i,j}  \;
\big(\mathcal{R}_{i2}+i\mathcal{R}_{i3}\big)
\big(\mathcal{R}_{j2}-i\mathcal{R}_{j3}\big) \; \mathcal{N}_{ij} \; ,
\label{sigmaqq}
\end{align}
with the reduced amplitudes
\be\label{eq:Nij}
\mathcal{N}_{ij} \;\equiv\;
  \frac{(\hat s-2m_q^2)\; \text{Re}\big(y_q^{\varphi_i^0}  y_q^{\varphi_j^0 *} \big) - 2m_q^2 \; \text{Re}\big(y_q^{\varphi_i^0}  y_q^{\varphi_j^0 }\big)}{(\hat s-M_{\varphi_i^0}^2 + i M_{\varphi_i^0} \Gamma_{\varphi_i^0})\, (\hat s-M_{\varphi_j^0}^2-iM_{\varphi_j^0}\Gamma_{\varphi_j^0})} \; .
\ee
We have kept the dependence on the initial quark masses, since otherwise the $q\bar q$ Yukawa coupling vanishes. This implies a strong suppression of this production mechanism by a factor $m_q^2/v^2$.

The gluon fusion mechanism dominates by far the previous one. The corresponding LO cross section at the partonic level takes the form
\begin{align}
 \hat{\sigma}(gg\to H^+ W^-)\; & =\; \frac{\alpha_s^2 \; T_F^2}{4096 \; \pi^3 \; v^4} \; \lambda^{3/2}(\hat s,M_{H^\pm}^2,M_W^2)
\notag \\
  & \qquad\qquad\qquad \times \; \sum_{i,j} \; (\mathcal{R}_{i2} + i \mathcal{R}_{i3})(\mathcal{R}_{j2} - i \mathcal{R}_{j3}) \;  \mathcal{G}_{ij} \; ,
\label{sigmagg}
\end{align}
where $T_F=1/2$ is the $SU(3)$ colour group factor and the reduced amplitudes $\mathcal{G}_{ij}$ are given by
\be\label{eq:Gij}
\mathcal{G}_{ij}\; \equiv\; \sum_{q q'}\;
 \frac{\text{Re}\big( y_q^{\varphi_i^0} \big)\, \text{Re}\big( y_{q'}^{\varphi_j^0} \big)  \,\mathcal{F}(x_q)\,\mathcal{F}(x_{q'})^*\, +\,
\text{Im}\big( y_q^{\varphi_i^0} \big)\, \text{Im} \big( y_{q'}^{\varphi_j^0} \big)\,
 \mathcal{K}(x_q)\,\mathcal{K}(x_{q'})^*
 }{ (\hat s-M_{\varphi_i^0}^2+ i M_{\varphi_i^0} \Gamma_{\varphi_i^0}) \, (\hat s-M_{\varphi_j^0}^2 - i M_{\varphi_j^0} \Gamma_{\varphi_j^0}) }\; ,
\ee
with $x_q\equiv 4m_q^2/\hat s$.
The explicit expressions of the different loop functions are:
\be
\mathcal{F}(x)\; =\; \frac{x}{2}\, [4+(x-1)f(x)] \, ,
\qquad\qquad\qquad
\mathcal{K}(x) \; = \;  - \frac{x}{2}\, f(x) \, ,
\ee
with
\be
f(x)\; =\; \begin{cases} \; -4\arcsin^2(1/\sqrt{x})\, , \qquad\quad & x\geqslant1
\\[3pt]\; \Big[\ln\Big( \frac{1+\sqrt{1-x}}{1-\sqrt{1-x}}\Big)- i\pi \Big]^2\, , & x<1 \end{cases} \; .
\ee
We have regulated the propagator poles with the term $ i M_{\varphi_i^0} \Gamma_{\varphi_i^0}$, both in Eqs.~(\ref{eq:Nij}) and (\ref{eq:Gij}),
because in our analysis one of the neutral scalars will, most likely, reach the on-shell kinematical region.
NLO QCD corrections to the gluon fusion channel are also available and will be taken into account; the details are given in appendix~\ref{QCDcorrections2}.


\section{Phenomenology}
\label{sec:phenom}

In the following phenomenological analysis, besides the fermiophobic charged-Higgs assumption ($\varsigma_f\to 0$),
we are also going to consider that the Higgs potential is CP-conserving. The consequence of this last hypothesis is that the CP-odd neutral Higgs $A$ will also be fermiophobic, as we have mentioned before in section \ref{sec:A2HDM};
moreover $\lambda_{AH^+H^-} =0$. This means that the decay $H^+\to  W^+ A^* \to W^+ \bar{f}f$ does not occur and $A$ does not contribute either to
$H^+\to W^+\gamma$. The charged-Higgs production amplitudes mediated by a virtual $A$
also vanish. The CP-odd scalar can contribute to $H^\pm$ production in a direct way through the $q_u \bar{q}_d \to W^{*} \to H^+ A$ production channel or, in an indirect way, by modifying the total decay rate $\Gamma_{\varphi_i^0}$,
which regulates the pole in the CP-even scalar propagators ($\varphi_i^0=h,\, H$), through
decays like $\varphi_i^0\to AA$ or $\varphi_i^0\to AZ$. The decay $H\to Ah$ cannot occur at tree level because all cubic vertices of the scalar potential involving an odd number of $A$ fields vanish in the CP-conserving limit. The total decay width $\Gamma_{\varphi_i^0}$ is the sum of all the decay rates explicitly presented in appendix~\ref{app:decayrates}.

In our particular case, the expressions for the Yukawa couplings simplify and become equal to the reduced scalar couplings to two gauge bosons. They are given by
\be
y^h_f = \frac{g_{hVV}^{\phantom{\mathrm{SM}}}}{g_{hVV}^{\mathrm{SM}}} = \mathcal{R}_{11} =  \cos\tilde\alpha \, ,
\qquad
y^H_f = \frac{g_{HVV}^{\phantom{\mathrm{SM}}}}{g_{hVV}^{\mathrm{SM}}} = R_{21} =  -\sin\tilde\alpha \, ,
\qquad
y^A_f = g_{AVV}^{\phantom{\mathrm{SM}}} = \mathcal{R}_{31} = 0 \, .
\ee

Even within the restricted range of charged-Higgs masses we are interested in, $M_{H^\pm} \in [M_W,M_W+M_Z]$, the possible phenomenological signals depend on the choice of masses for the remaining scalars. In the following subsections, we will therefore consider different scenarios for the scalar spectrum. The first part of the analysis will be dedicated to the study of the various decay modes of the charged Higgs and the second part will focus on estimating the production cross sections.


\subsection{Decay rates and branching ratios}

One of the two CP-even scalars should correspond to the Higgs boson discovered at the LHC, but a broad range of masses is allowed for the other two neutral scalars. We will consider the following four scenarios, which cover the different possibilities:
\begin{enumerate}
\item $M_h = 125$~GeV \ and \ $M_{A,H} > M_W + M_Z$.
\item $M_h = 125$~GeV \ and \  $M_{A} < M_W + M_Z < M_{H}$.
\item $M_h = 125~\mathrm{GeV} < M_H < M_W + M_Z$ \ and three different options for $A$
($M_A < M_H$, \ $M_H < M_A < M_W + M_Z$ \ and \ $M_A > M_W + M_Z$).
\item $M_H = 125$~GeV, \ $M_h = 90$~GeV  \ and \  $M_{A} < M_W + M_Z$.
\end{enumerate}

\subsubsection{First Scenario}
\label{sec:phenom1}

\begin{figure}[t]
\centering
\includegraphics[scale=0.53]{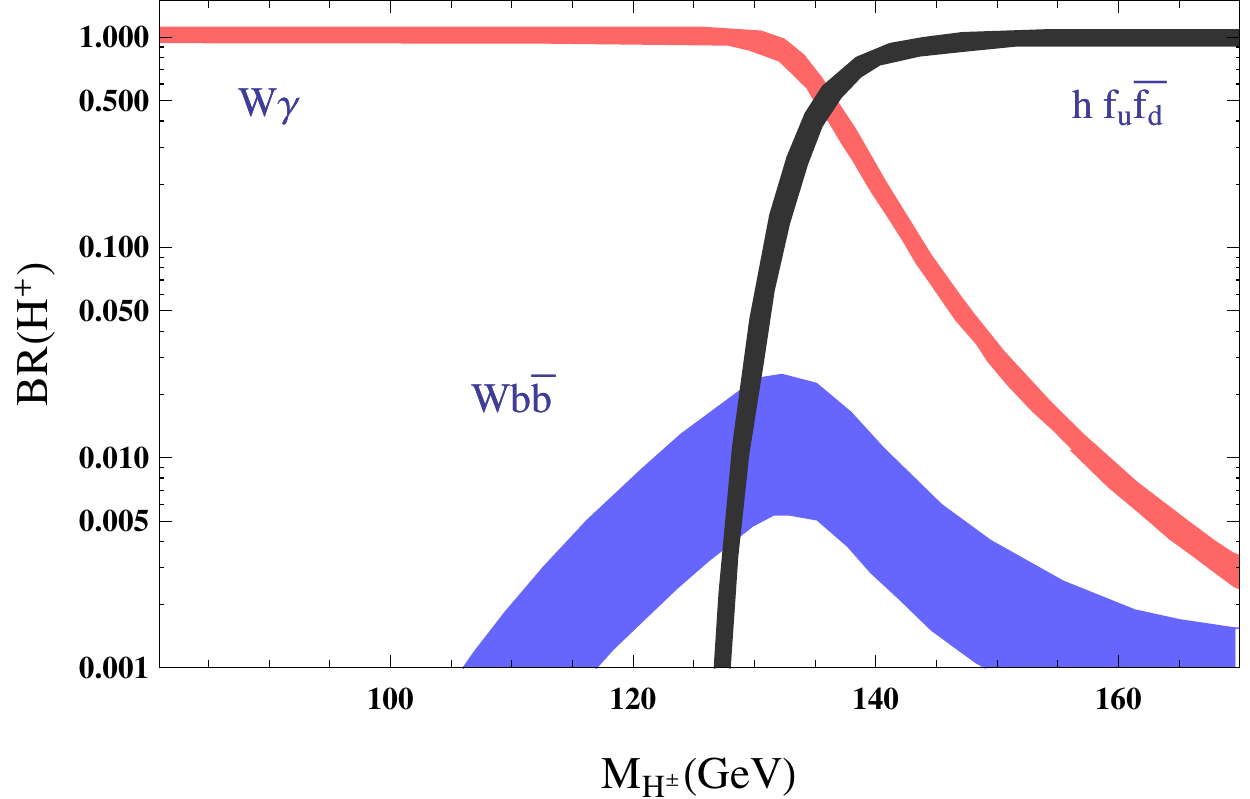} \;\;\; \includegraphics[scale=0.53]{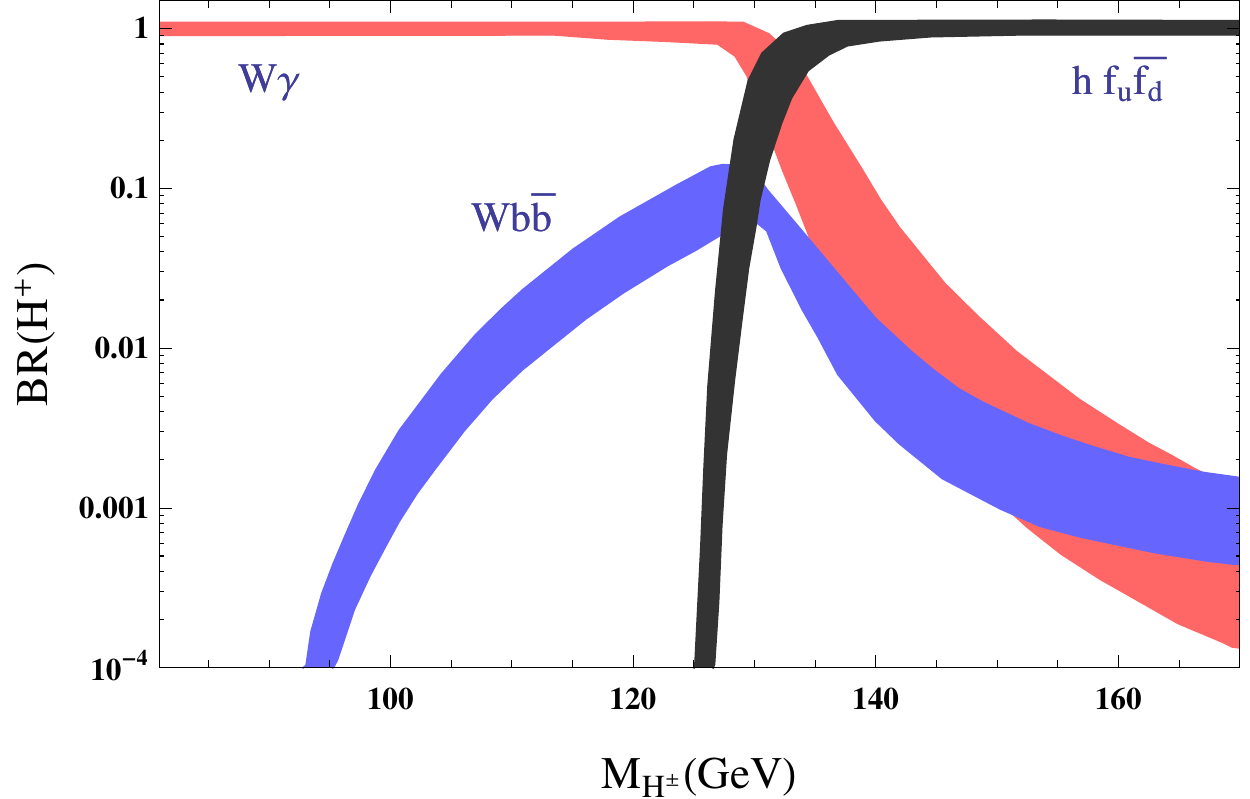} \\[2ex]
\includegraphics[scale=0.53]{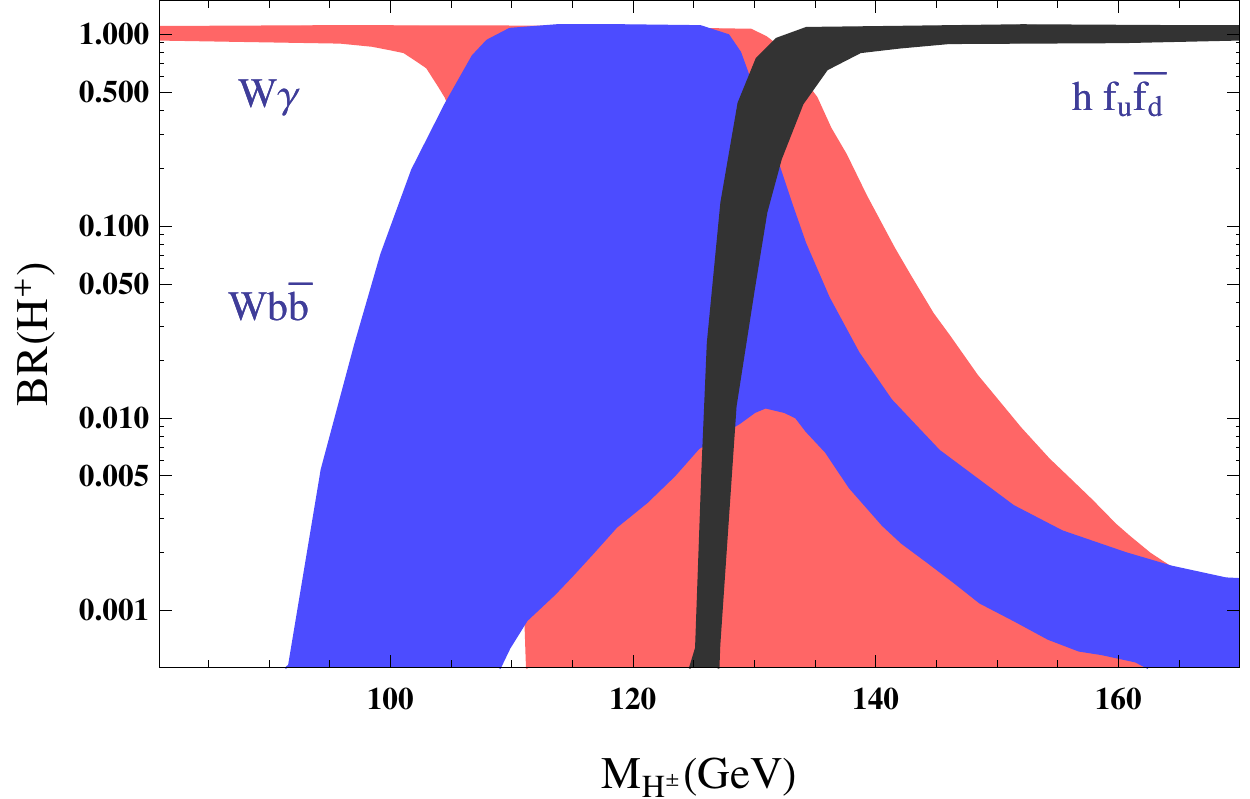} \;\;\; \includegraphics[scale=0.55]{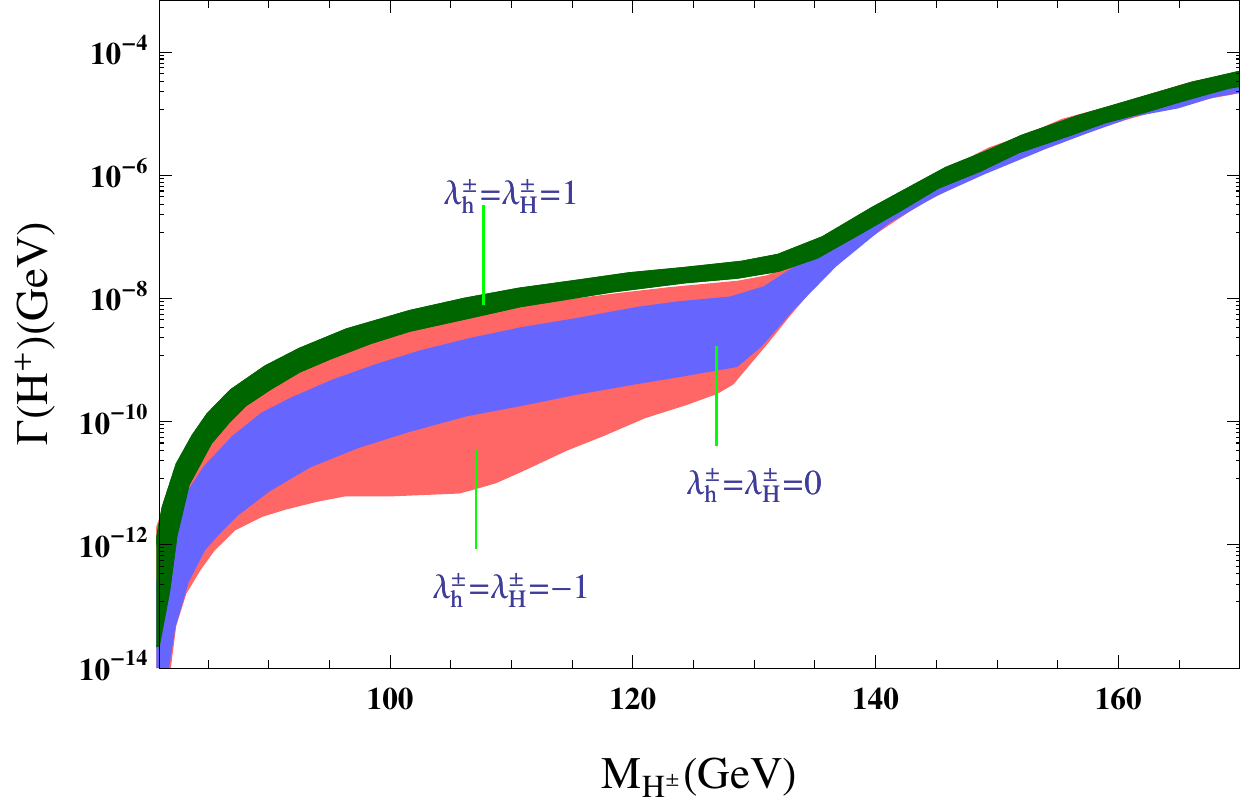}
\caption{Charged-Higgs branching ratios as functions of $M_{H^\pm} \in [M_W,M_W+M_Z]$, for $\cos\tilde\alpha = 0.9 $, $M_H \in [M_W+M_Z, \, 500\, \text{GeV}]$ and $\lambda_{h H^+ H^-} = \lambda_{H H^+ H^-} = 1$ (top-left), 0 (top-right) and -1 (bottom-left). The corresponding total decay widths are shown in the bottom-right panel $(\lambda_{h}^{\pm} \equiv \lambda_{h H^+ H^-} ,\; \lambda_{H}^\pm  \equiv \lambda_{H H^+ H^-})$.}
\label{11C09}
\end{figure}

In the first scenario the mass of the lightest CP-even scalar is set to $M_h=125$ GeV. Therefore, the strong constraint on the scalar mixing angle, from the global fit to the light Higgs boson signal strengths using the LHC
data, must be used: $|\cos\tilde\alpha|> 0.9$ at 68\% CL \cite{ilisie3,ilisie2,ilisie1}.
The masses of the remaining neutral scalars are considered to be greater than $M_W + M_Z$ so that decays of a charged Higgs into an on-shell $H$ or $A$ are kinematically forbidden. In the limit $\cos\tilde\alpha \to 1$, the only surviving decay amplitude (not proportional to $\sin\tilde\alpha$) is the contribution of $H$ to the amplitude $S_{(2)}$. Thus, in this limit the branching ratio of $H^+\to W^+ \gamma$ is 100\%; all the other decay channels vanish.

If we set $\cos\tilde\alpha = 0.9$, $\lambda_{h H^+ H^-} = \lambda_{H H^+ H^-}= 1$, vary the charged Higgs mass in the region $M_{H^\pm} \in [M_W,M_W+M_Z]$ and $M_H$ from $M_W + M_Z$ up to 500 GeV, we obtain the branching ratios (top-left) and total decay width (bottom-right) shown in Fig.~\ref{11C09}. The width of the branching ratio bands reflects the variation of the input parameters in the mentioned ranges. The same consideration is valid for the following scenarios. The decay channel $H^+\to W^+ \gamma$ dominates for $M_{H^{\pm}} \lesssim M_h$. When the charged Higgs is kinematically allowed to decay into an on-shell $h$, then $H^+\to hf_u\bar{f}_d$ rapidly becomes the dominant channel as $M_{H^\pm}$ grows. The remaining $H^+\to W^+ b\bar b$ branching ratio stays at a few percent level or less for the whole allowed region. The total decay width approximately grows from $10^{-14}$ up to $10^{-8}$ GeV, in the region dominated by the radiative $H^+\to W^+ \gamma$ decay, and sizeably increases up to $10^{-5}$ GeV, once the $hf_u\bar{f}_d$ production threshold is reached.
The tree-level decay rates are significantly larger than the loop-induced one. Flipping the sign of $\cos\ta$ leads to an equivalent solution with a sign flip of the coupling $\lambda_{hH^+H^-}$. This is also valid for the next scenarios.

If, instead, we consider all the previous settings but taking this time $\lambda_{h H^+ H^-} = \lambda_{H H^+ H^-}= 0$, then the only amplitude that contributes to the $H^+\to W^+\gamma$ decay channel is $S_{(3)}$, which is suppressed by a factor $\sin{\tilde\alpha}$. As shown in Fig.~\ref{11C09} (top-right), this channel remains the dominant one up to $M_{H^\pm} \gtrsim M_h$, but with a sizeably smaller decay width
(bottom-right). The $H^+\to W^+ b\bar b$ branching ratio is also more sizeable, raising up to the 10\% level.

Let us now consider $\lambda_{h H^+ H^-} = \lambda_{H H^+ H^-}  = -1$ and everything else as previously. In this particular case the amplitudes $S_{(2)}$ and $S_{(3)}$ interfere destructively and, as a consequence, the decay $H^+\to W^+b\bar{b}$ competes with $H^+\to W^+\gamma$. Thus, the $Wb\bar{b}$ decay channel can dominate in some cases. However, as soon as the charged Higgs reaches  $M_{H^\pm} \gtrsim M_h$, the dominant decay mode is again $H^+\to hf_u\bar{f}_d$, as in the previous cases (Fig.~\ref{11C09}, bottom-left).


\subsubsection{Second Scenario}
\label{sec:phenom2}

In the second scenario the mass of lightest CP-even scalar is set to $M_h=125$ GeV
and $M_H > M_W+M_Z$, as in the first one, but this time we assume the CP-odd Higgs boson $A$ to have its mass below the $WZ$ threshold ($M_A < M_W+M_Z$).
The decay of the charged Higgs into an on-shell $A$ is then kinematically allowed, but into an on-shell $H$ is forbidden. The same constraint as before is considered for the scalar mixing angle. Taking the limit $\cos\tilde\alpha \to 1$, this time there are two surviving decay amplitudes, $H^+\to W^+ \gamma$ and $H^+\to A f_u \bar{f}_d$.

\begin{figure}[t]
\centering
\includegraphics[scale=0.53]{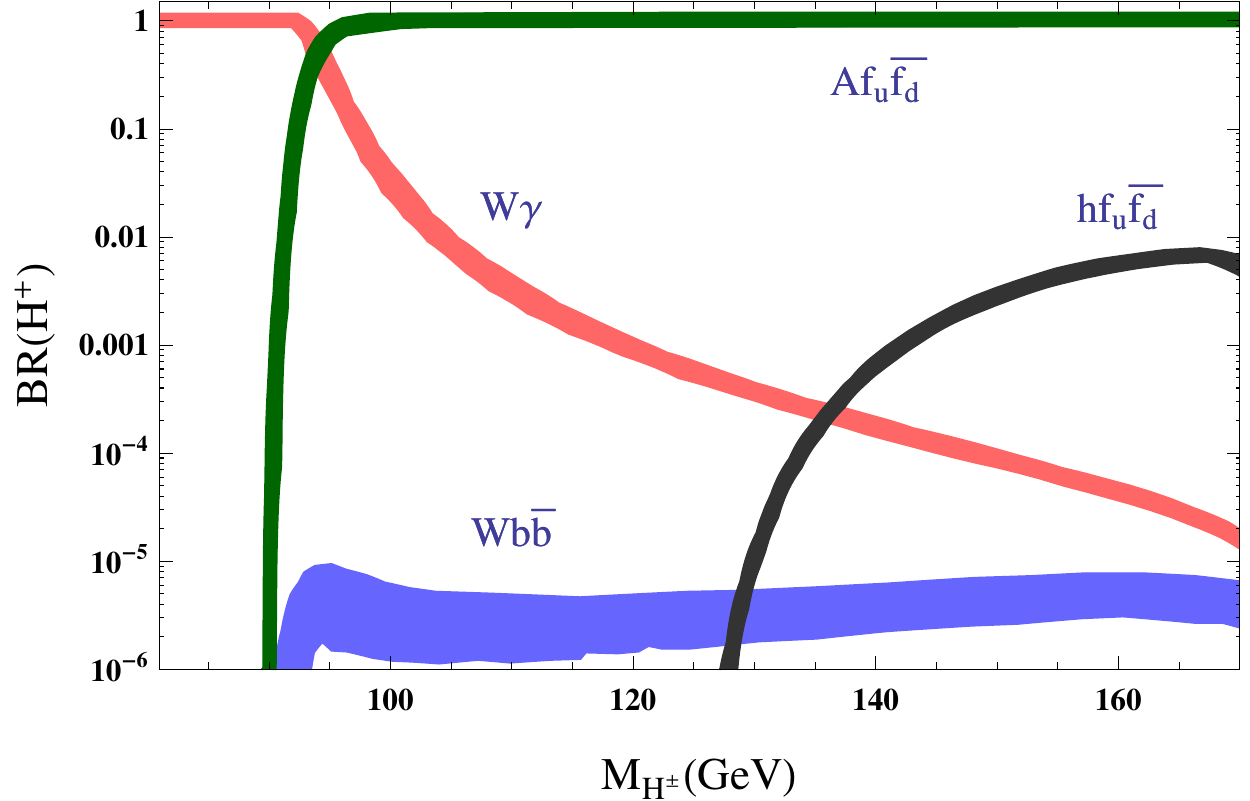} \;\;\; \includegraphics[scale=0.53]{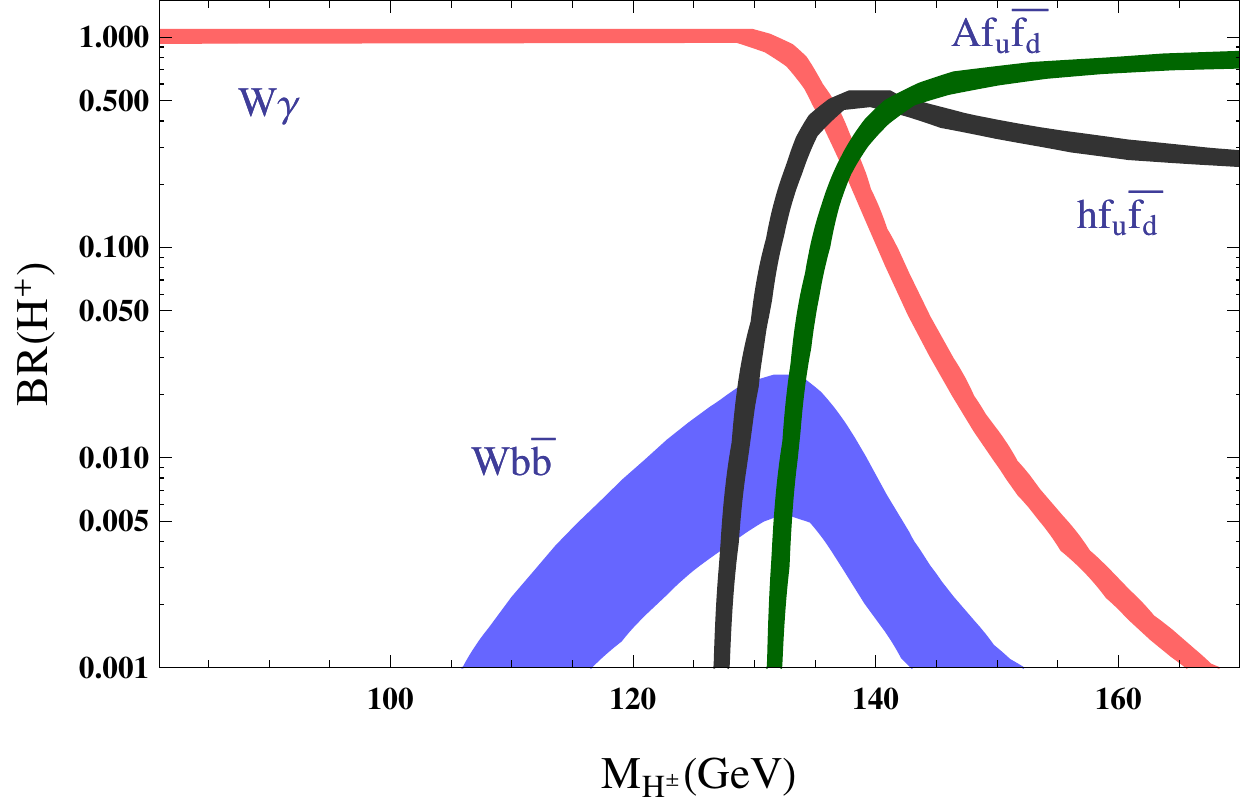} \\[2ex]
\includegraphics[scale=0.53]{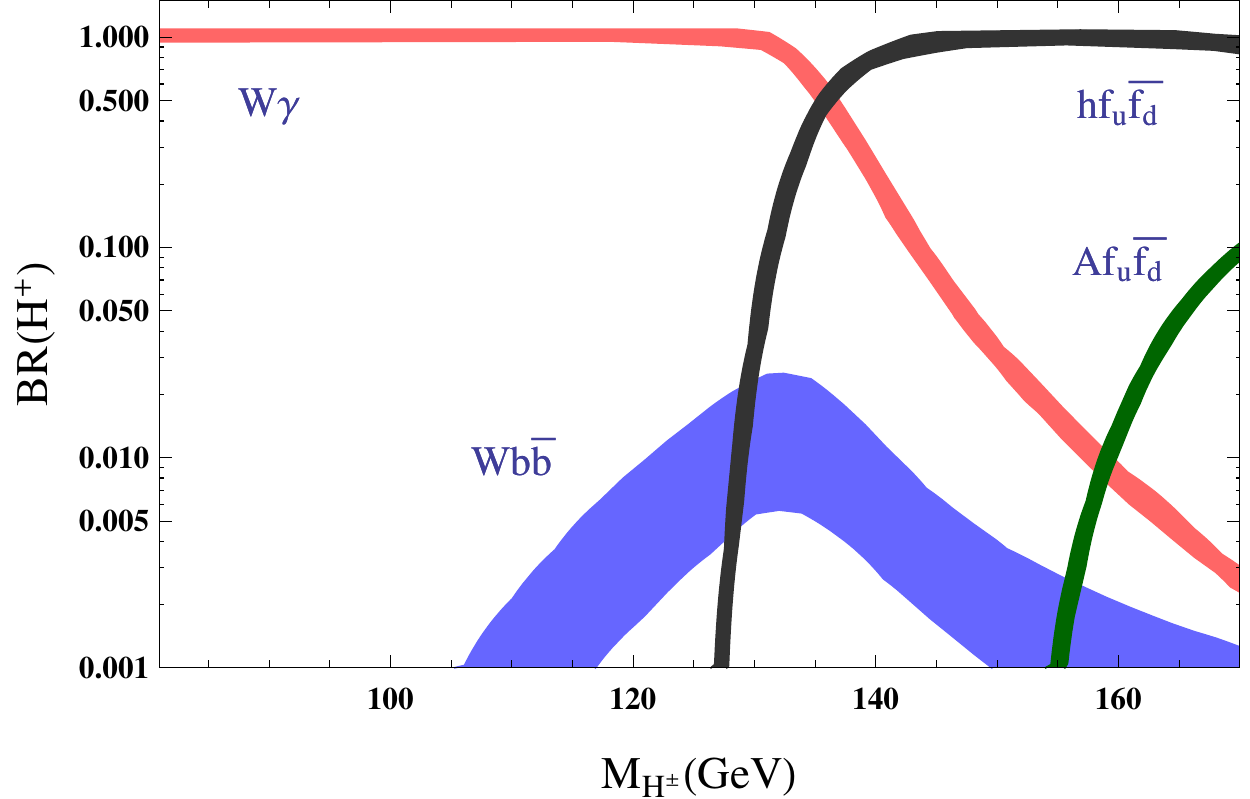} \;\;\; \includegraphics[scale=0.55]{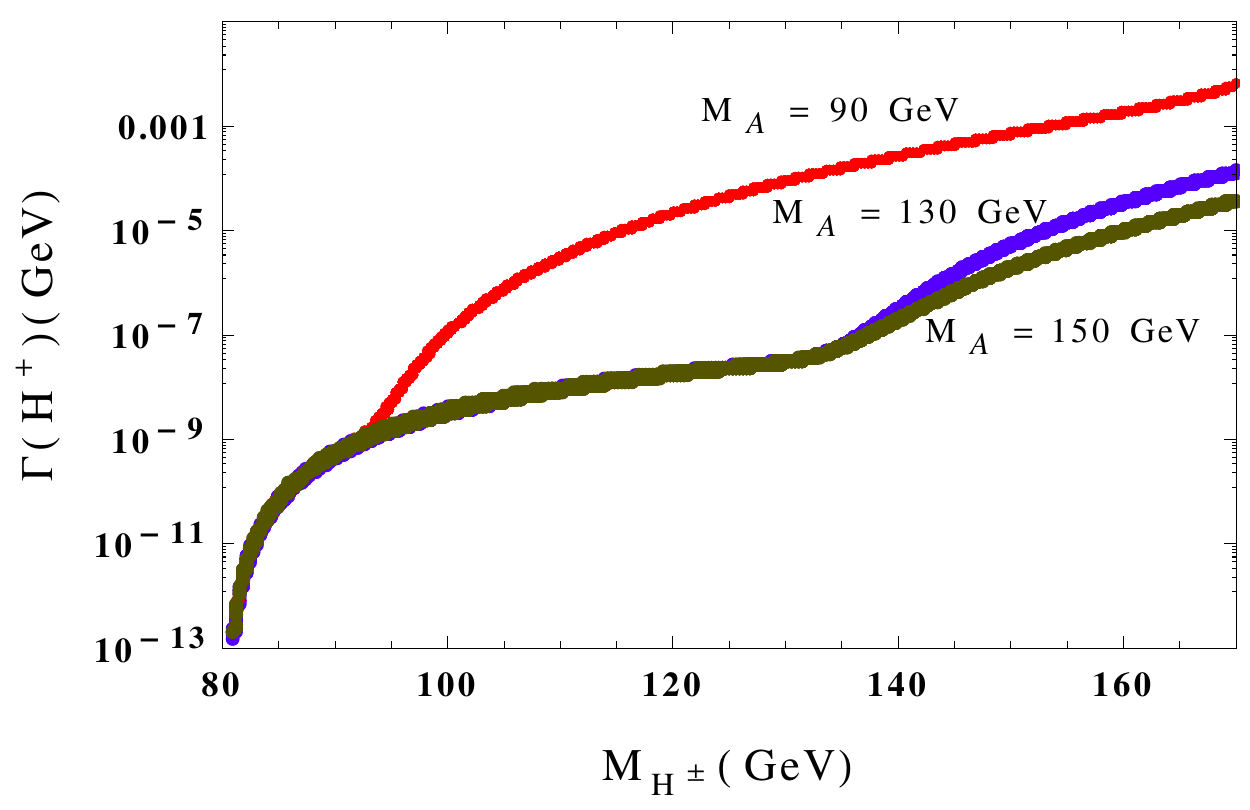}
\caption{\it Charged-Higgs branching ratios as functions of $M_{H^{\pm}}$, for $\lambda_{h H^+ H^-} = \lambda_{H H^+ H^-}= 1$, $\cos\tilde\alpha=0.9$ and $M_A = $ 90 (top-left), 130 (top-right) and 150 (bottom-left) GeV. $M_H$ is varied from $M_W + M_Z$ up to its permitted value by the oblique parameters. The bottom-right panel shows the corresponding total decay widths.}
\label{11C09A90130150}
\end{figure}

Let us consider $\cos\tilde\alpha=0.9$, $\lambda_{hH^+H^-}=\lambda_{HH^+H^-}=1$ and $M_A=$ 90, 130 and 150 GeV. For each value we shall vary $M_H$ from $M_W+M_Z$ up to its allowed upper bound from the oblique parameters (at 68\% CL) \cite{ilisie3,ilisie2,ilisie1}, with a maximum limit of 500 GeV. We obtain then the branching ratios and total decay widths in Fig.~\ref{11C09A90130150}. We observe that for $M_A=90$ GeV (top-left), when kinematically allowed, the decay to an on-shell $A$ boson rapidly becomes the dominant one as $M_{H^\pm}$ increases. For this configuration the $Wb\bar{b}$ channel is insignificant. When $M_A=130$ GeV (top-right), which is close to $M_h$, the decays into an on-shell $h$ or $A$ boson compete. However, the decay to $Af_u\bar{f}_d$ still dominates even if the masses are similar because of the relative suppression factor $\sin^2\tilde\alpha$ of the $hf_u\bar{f}_d$ width. As $M_A$ becomes heavier, $M_A=150$~GeV (bottom-left), the decay rate into an on-shell $A$ boson does not grow as
rapidly as in the previous cases; thus, $hf_u\bar{f}_d$ dominates over $Af_u\bar{f}_d$ in the considered region. For the last two configurations, that is $M_A=130$ and 150 GeV, the $H^+\to W^+b\bar{b}$ decay channel can also bring sizeable contributions.

The total decay width in this scenario can reach as high as $10^{-3}$ GeV, see Fig.~\ref{11C09A90130150} (bottom-right). This is approximately two orders of magnitude larger than in the previous case and it is due to the tree-level decays, as we mentioned earlier. The maximum values are reached for the smallest mass of the CP-odd scalar ($M_A=90$ GeV).

It is worth mentioning that, just as in the previous scenario, the $Wb\bar{b}$ branching ratio can be sizeably increased by decreasing the $W\gamma$ decay width through a sign flip of the $\lambda_{\varphi_i^0H^+H^-}$ couplings, creating destructive interference among the various loop contributions. The same consideration is also valid for the next scenario.

\subsubsection{Third Scenario}
\label{sec:phenom3}

\begin{figure}[t]
\centering
\includegraphics[scale=0.53]{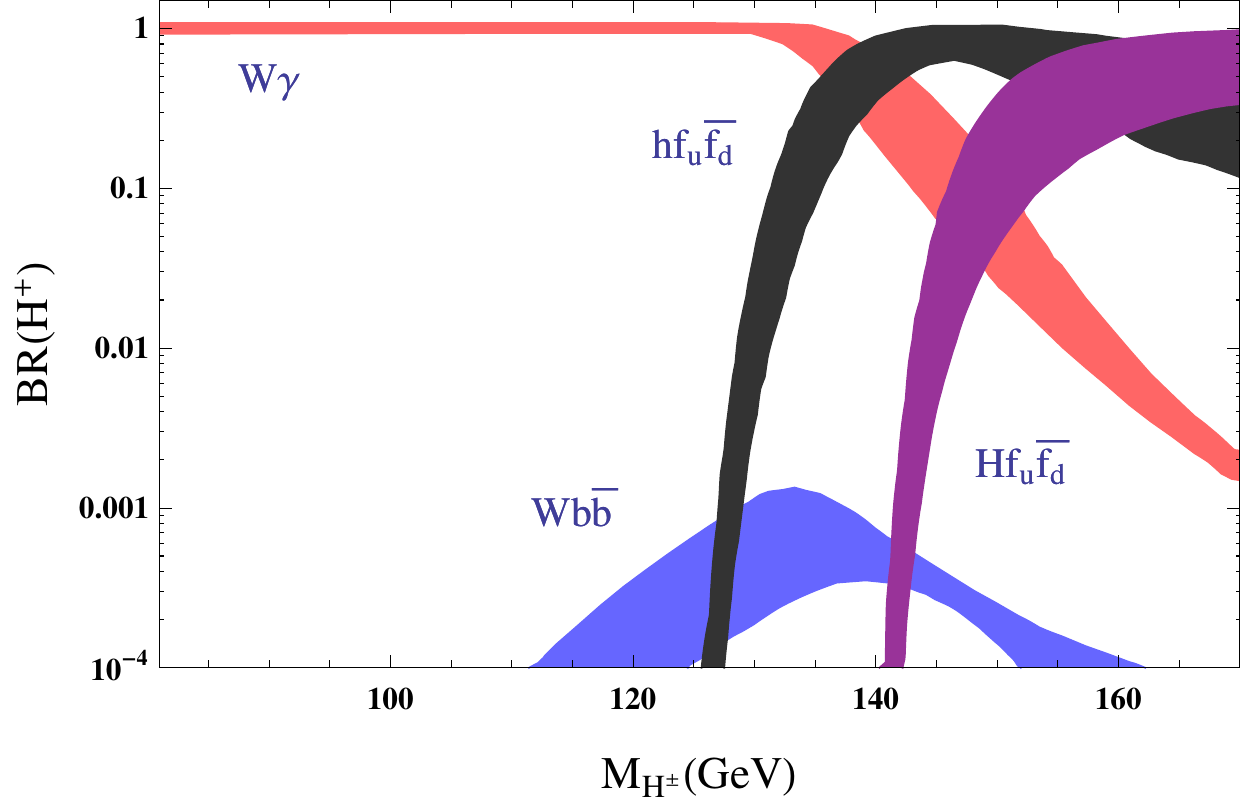} \;\;\; \includegraphics[scale=0.53]{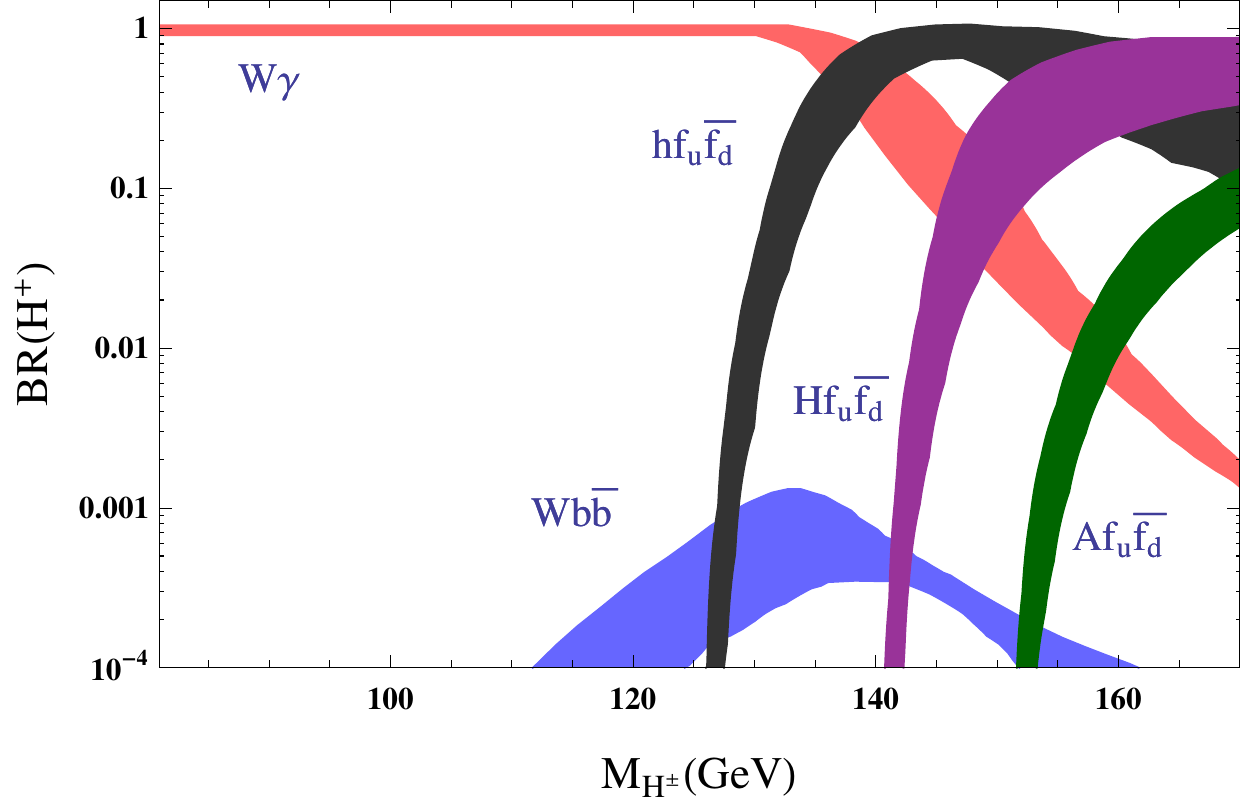} \\[2ex]
\includegraphics[scale=0.53]{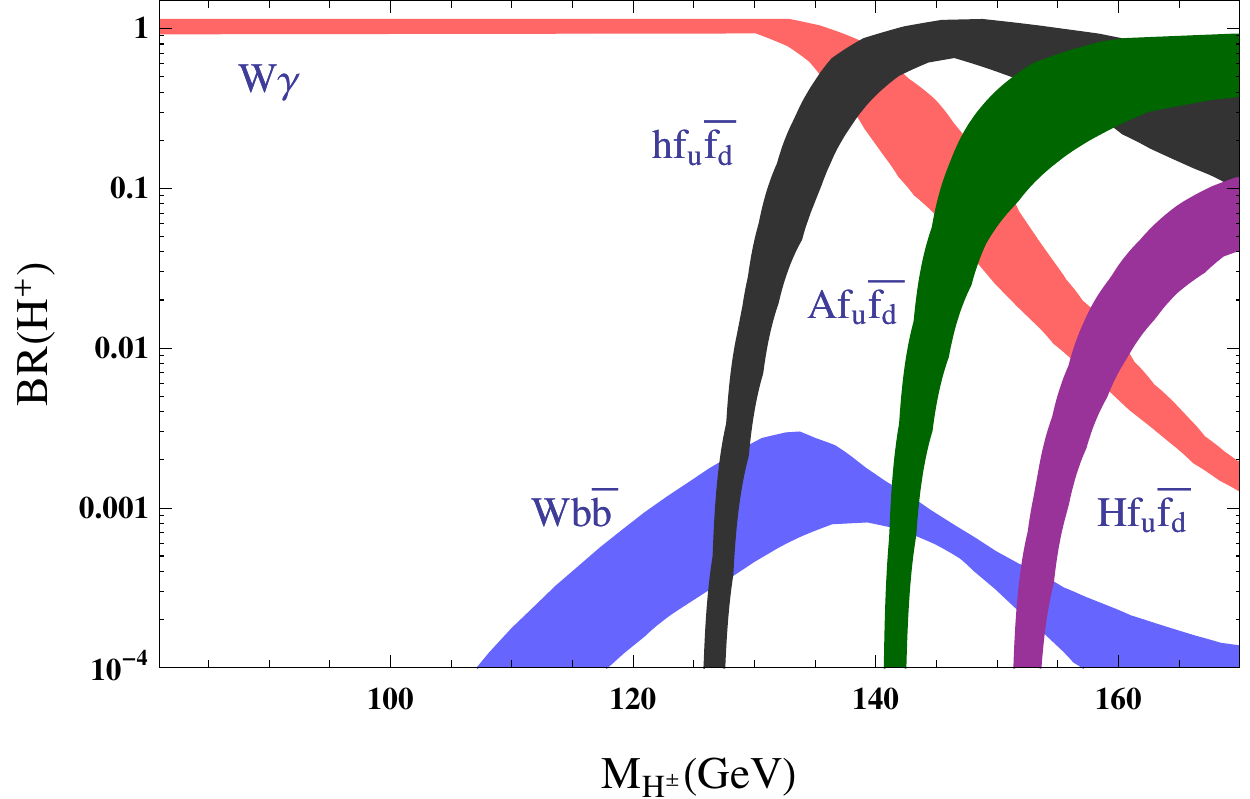} \;\;\; \includegraphics[scale=0.55]{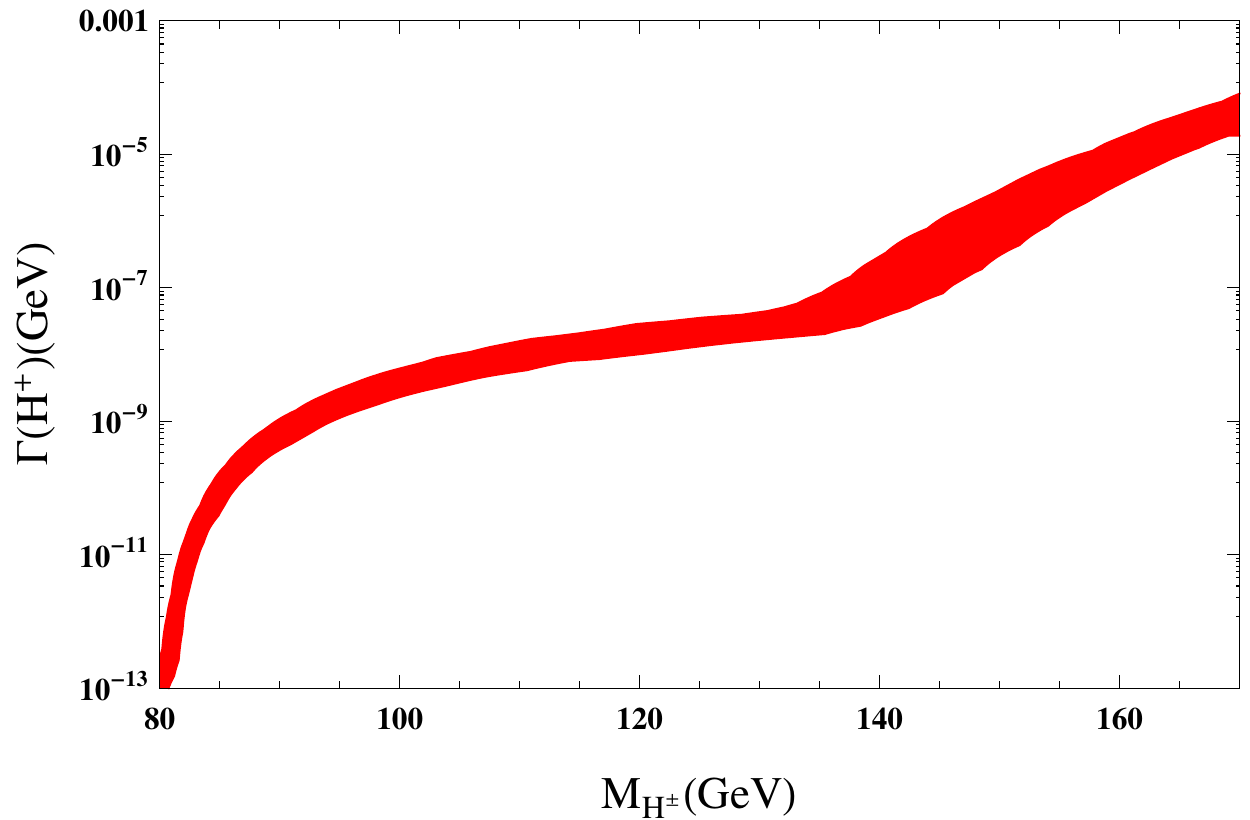}
\caption{\it Charged-Higgs branching ratios as functions of $M_{H^{\pm}}$, for $\lambda_{h H^+ H^-} = \lambda_{H H^+ H^-}= 1$, $\cos\tilde\alpha\in[0.9,\, 0.99]$,  $M_H = 140$ GeV, $M_A > M_W + M_Z $ (top-left); $(M_H, M_A)=(140, 150)$ GeV (top-right) and $(M_H, M_A)=(150, 140)$ GeV (bottom-left). The total decay width for the first case is also shown (bottom-right).}
\label{case3}
\end{figure}

In this scenario the mass of the lightest CP-even scalar is also set to $M_h=125$ GeV, while the heavy CP-even Higgs boson $H$ has its mass in the range $M_h < M_H < M_W+M_Z$. For the mass of the remaining CP-odd scalar we consider three different possibilities: a) $M_A > M_W+M_Z$, so that the decay into an on-shell $A$ is forbidden; b) $M_H < M_A < M_W+M_Z$, and c) $M_A < M_H < M_W + M_Z$. In the last two situations the $H^\pm$ boson could decay into any of the three neutral scalars. Again, we use the LHC constraint $|\cos\tilde\alpha|>0.9$ at 68\% CL. In the limit $\cos\tilde\alpha \to 1$, there are three possible surviving decay channels: $H^+\to W^+ \gamma$, $H^+\to H f_u \bar{f}_d$ and, when kinematically allowed, $H^+\to A f_u \bar{f}_d$.

For all three cases we set $\lambda_{h H^+ H^-} = \lambda_{H H^+ H^-}= 1$ and vary $\cos\tilde{\alpha}\in [0.9,\, 0.99]$.
In Fig.~\ref{case3} we show the $H^\pm$ branching ratios (top-left) and total decay width (bottom-right) when $M_H$ = 140 GeV and $M_A > M_W + M_Z$ (first case). To illustrate the other two possibilities, we set $(M_H, M_A)=(140, 150)$ GeV (Fig.~\ref{case3}, top-right) and $(M_H, M_A)=(150, 140)$ GeV (Fig.~\ref{case3}, bottom-left). The total $H^\pm$ decay widths for these two last configurations are very similar to the first one.

The $H^\pm$ decay into an on-shell $h$ boson has a global relative suppression factor of $\tan^2{\tilde\alpha}$ with respect to the decay into an on-shell $H$ and $\sin^2{\tilde\alpha}$  with respect to the decay into an on-shell $A$. Therefore, when $hf_u\bar{f}_d$ competes with $Hf_u\bar{f}_d$, the later one dominates as $\cos\tilde\alpha\to 0.99$ (Fig.~\ref{case3}, upper-left). When all three channels compete, the decay rate into the heaviest scalar boson grows the slowest and, therefore, brings a sub-dominant contribution to the branching ratios.


\subsubsection{Fourth Scenario}
\label{sec:phenom4}

In this last scenario we are going to set the mass of the heavy CP-even scalar to $M_H=125$ GeV; therefore, the LHC bounds
translate into $|\sin\tilde\alpha|>0.9$ at 68\% CL. The mass of the light CP-even scalar will be set to $M_h=90$ GeV. As for the CP-odd one, we will consider three possible values: $M_A$ = 150, 140 and 110 GeV.

In order to safely avoid the stringent constraints on light scalar masses from LEP \cite{Abbiendi:2002qp,Barate:2003sz}, we need to have very suppressed decay and production channels.  In our particular case with $\varsigma_f=0$, CP-conserving potential, and $M_A>M_h$ (therefore the decays $h\to AA$ and $h\to AZ$ are forbidden), we have the simple relation $\Gamma_h =\cos^2\tilde\alpha \; \Gamma^{\text{SM}}_h$. Here $\Gamma_h$ is the total decay rate of the light CP-even scalar boson with $M_h<M_H=125$ GeV, and $\Gamma^{\text{SM}}_h$ the corresponding decay rate in the SM for a Higgs boson with the same mass $M_h$.
The $\cos^2{\tilde\alpha}$ suppression factor is common to all allowed $h\to f\bar f$ decay modes, and cancels out in the branching ratios. The same suppression factor appears
in the LEP production rate, so that the signal strengths, relative to the SM, are then given by
\begin{align}
\mu_{X}^h \;\equiv\; \frac{\sigma(e^+e^-\to Zh) \; \text{Br}(h\to X)}{\sigma(e^+e^-\to Zh)_{\text{SM}} \; \text{Br}(h\to X)_{\text{SM}}}\; = \;\cos^2{\tilde\alpha} \; ,
\label{SignalStrX}
\end{align}
with $X=b\bar b$ and $\tau^+\tau^-$. Thus, we have a global suppression factor
$\cos^2{\tilde\alpha}$. The LEP constraints from the $\tau^+\tau^-$ channel, which are the strongest ones in our case, can then be avoided by setting $\cos^2{\tilde\alpha}\approx 0.02$ ($\sin\tilde\alpha\approx0.99$).
The OPAL collaboration has also performed a decay-mode-independent search for a light neutral scalar and found the upper limits $\cos^2\tilde{\alpha}<0.1$ (1) for $M_h < 19$ (81) GeV \cite{Abbiendi:2002qp}, which are weaker (in our case).

It is worth mentioning that in (\ref{SignalStrX}) we have ignored the charged-Higgs contribution to the $h\to\gamma\gamma$ decay rate. If however, we choose to enhance it
through the $H^\pm$ loop contribution,
it would only further suppress the fermionic branching ratios, weakening the bound on $\sin\tilde\alpha$.

\begin{figure}[t!]
\centering
\includegraphics[scale=0.53]{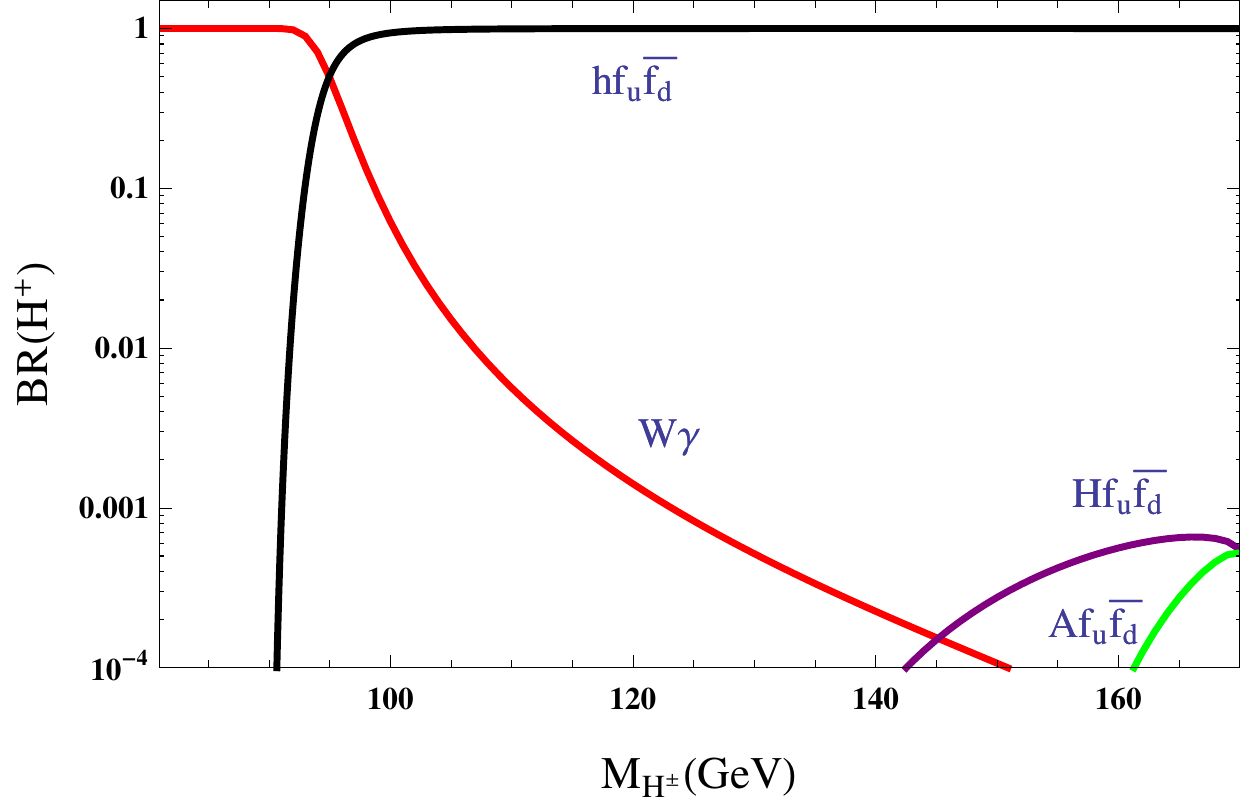} \;\;\; \includegraphics[scale=0.53]{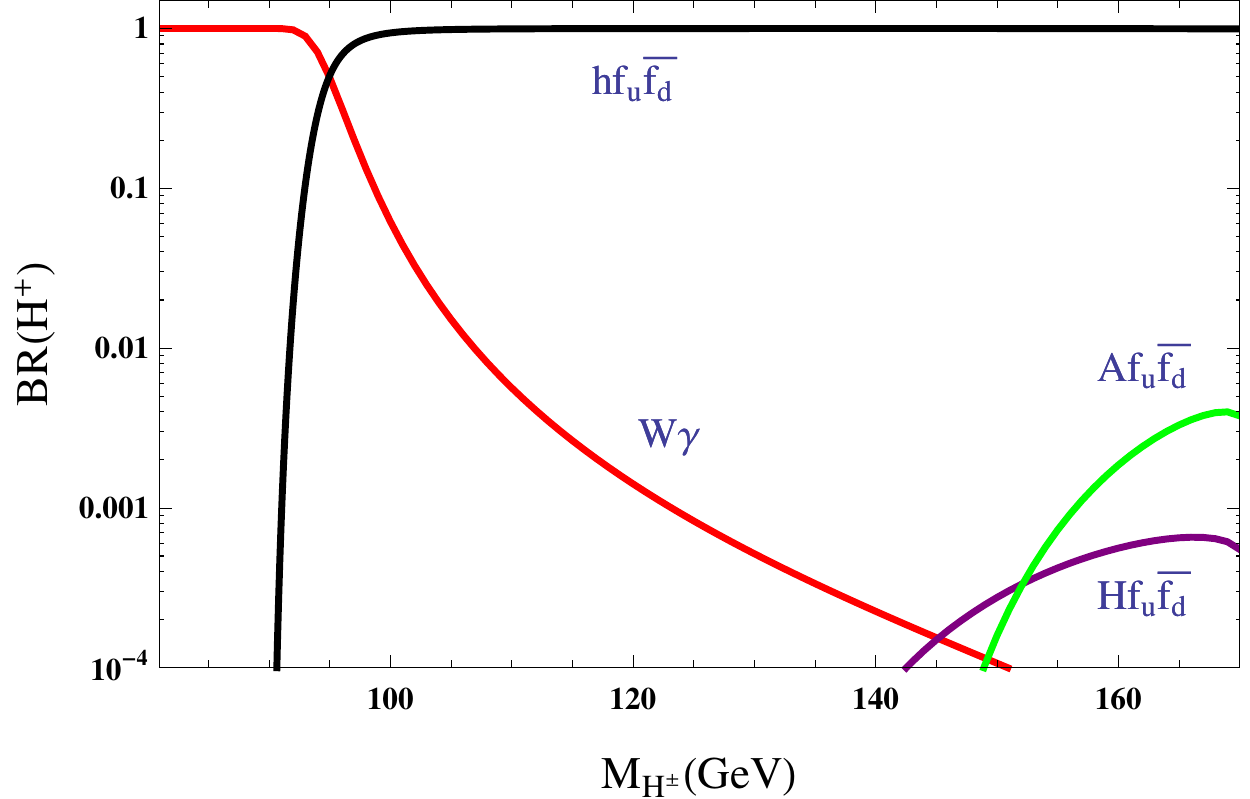} \\[2ex]
\includegraphics[scale=0.53]{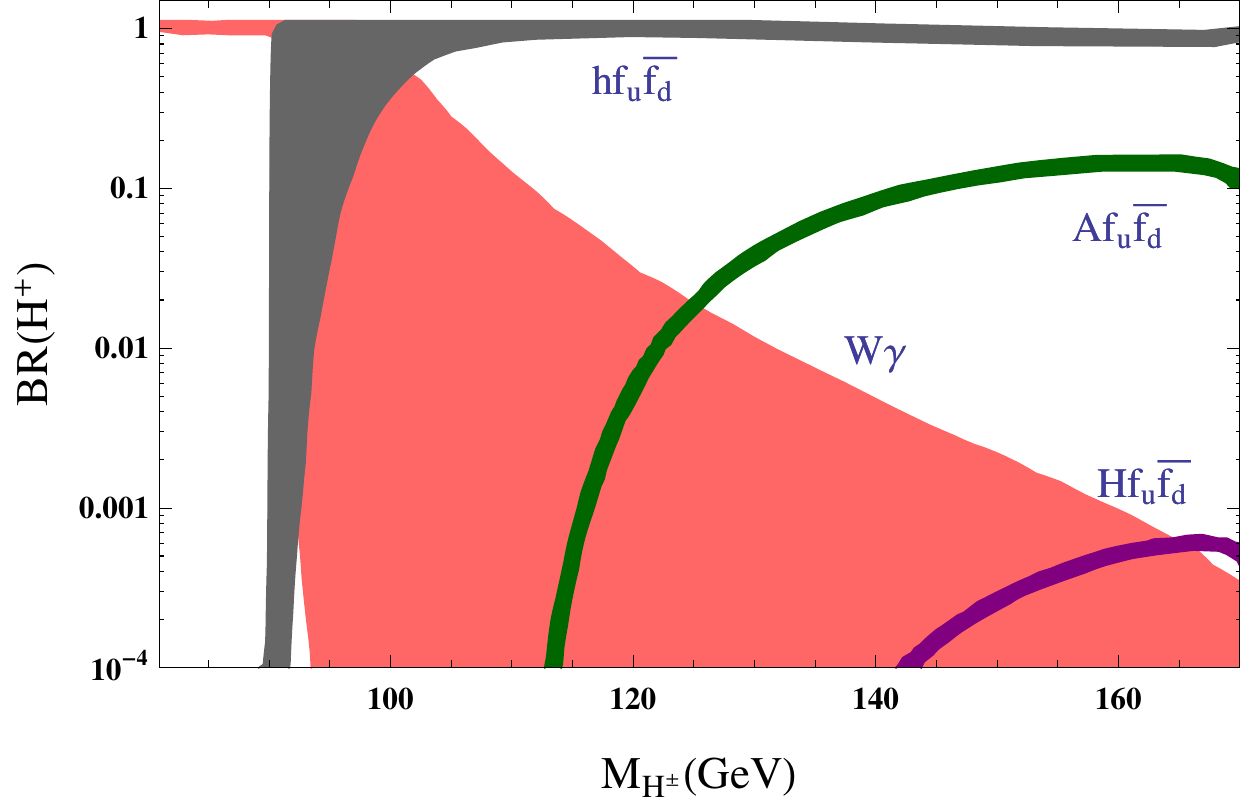} \;\;\; \includegraphics[scale=0.55]{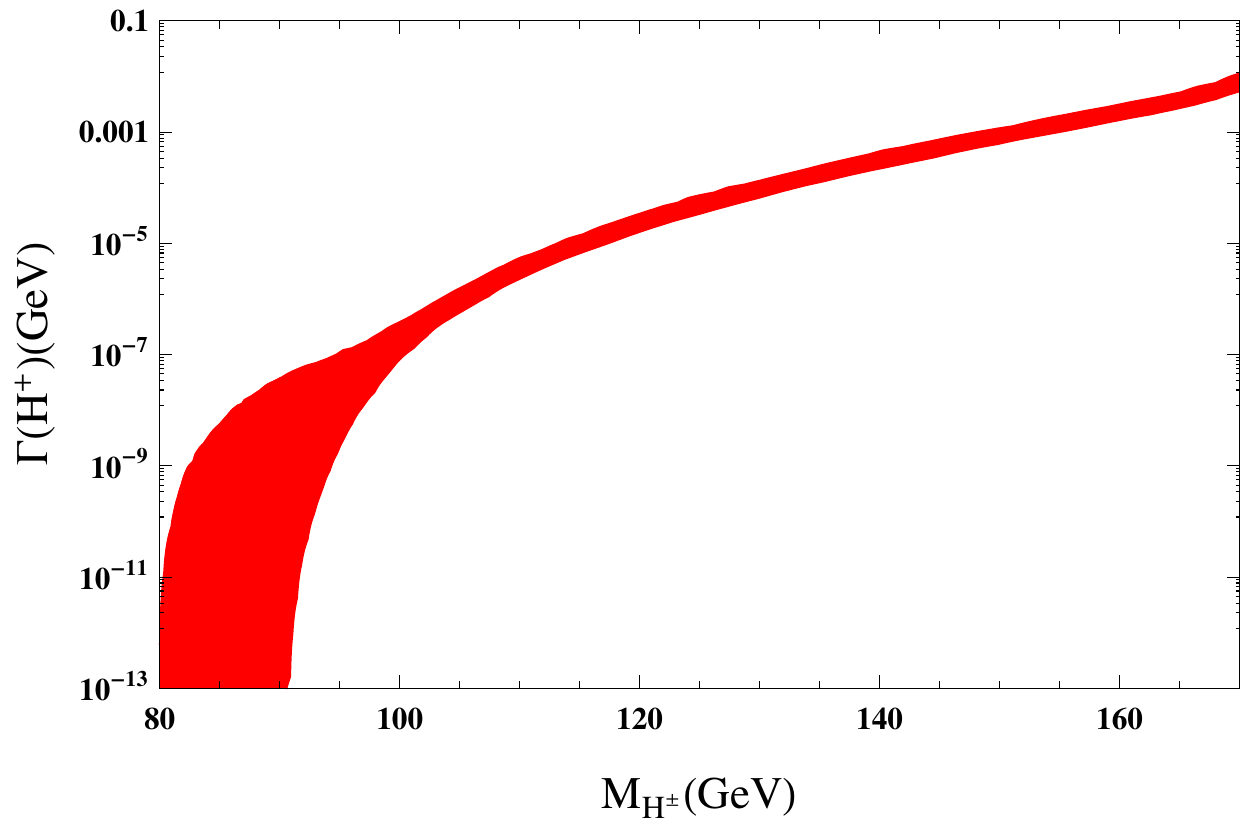}
\caption{\it Charged-Higgs branching ratios as functions of $M_{H^{\pm}}$, for $\sin\tilde\alpha = 0.99$ and $M_h = 90$ GeV.
The trilinear couplings are set to $\lambda_{h H^+ H^-} = \lambda_{H H^+ H^-}= 1$, with $M_A = 150 $ GeV (top-left) and $M_A = 140 $ GeV (top-right),
and $\lambda_{h H^+ H^-}, \lambda_{H H^+ H^-}\in[-5,5]$ with $M_A = 110 $ GeV (bottom-left). The total decay width (bottom-right) for the last case is also shown.}
\label{case4}
\end{figure}

With all this being said, we set $\sin\tilde\alpha=0.99$. In Fig.~\ref{case4} we plot the $H^\pm$ branching ratios for $M_A=150$ (top-left) and 140 GeV (top-right), taking $\lambda_{hH^+H^-}=\lambda_{HH^+H^-}$. In both plots we can observe that, when kinematically allowed, the tree-level $H^+\to hf_u\bar{f}_d$ decay dominates. In this case, this decay no longer has a suppression factor as its partial width is proportional to $\sin^2\tilde\alpha\sim 1$. The suppression factor appears now in the $Hf_u\bar{f}_d$ decay mode with a partial decay width proportional to $\cos^2\tilde\alpha$. This is why, when $M_A\sim M_H$, the decay into an on-shell $A$ boson dominates over the decay into an on-shell $H$.
Both $A$ and $H$ contributions are, however, very suppressed due to their heavy masses. It is also worth mentioning that a small variation of $M_A$ can produce a significant change (roughly, one
order of magnitude) in $\text{Br}(H^+ \to A f_u \bar{f}_d )$, as can be seen in Fig.~\ref{case4} (top-left and top-right).

For the last case we set $M_A$ to 110 GeV. The perturbativity bounds on neutral scalar couplings to a pair of charged Higgses, for the considered region of the charged Higgs mass, are roughly given by $|\lambda_{\varphi_i^0H^+H^-}|\leq5$ (here $\varphi_i^0=h,H$) \cite{ilisie1}. In order to see the impact of these two parameters on the $H^\pm$ branching ratios, we will vary both independently in this region.  The result, shown in Fig.~\ref{case4} (bottom-left), is that $W\gamma$ and $hf_u\bar{f}_d$ compete, even after crossing the $h$ production threshold. Since $M_A$ is lighter than in the previous two cases, the $H^+\to Af_u\bar{f}_d$ branching ratio can also reach higher values. The total decay rate for this configuration is also shown in Fig.~\ref{case4} (bottom-right).

As we have seen, in the four proposed scenarios, the configuration of the $H^\pm$ branching ratios depends very sensitively on the chosen parameters. However, we can draw some important conclusions. There are only a few decay channels to be analysed and the largest decay widths are the tree-level ones, corresponding to the on-shell production of scalar bosons. Thus, the number of decay channels decreases as the number of neutral scalar bosons that are heavier than the charged Higgs ({\it i.e.}, $M_{\varphi_i^0}>M_{H^\pm}$) increases. The $W\gamma$ decay mode can bring sizeable contributions below and close to the the on-shell production threshold of a scalar boson. Short after this threshold is reached, as $M_{H^{\pm}}$ grows, the $H^+\to W^+\gamma$ branching ratio rapidly decreases. As we have shown, the $H^+\to W^+b\bar b$ decay can dominate over $H^+\to W^+\gamma$ in some cases, depending on the values of the $\lambda_{\varphi_i^0H^+H^-}$ couplings. If a fermiophobic charged Higgs is finally discovered in this
mass range, the precise values of its mass and branching ratios would provide priceless information about all other parameters. The masses of the remaining scalars would also be highly constrained by the electroweak oblique parameters. These constraints were used in our second scenario, because they put an upper bound on $M_H$; we did not mention them in the other cases, since they do not bring additional constraints. The mean lifetime of a fermiophobic charged scalar is short, ranging from $10^{-11}$ to $10^{-23}$ s, making its direct detection very compelling at the LHC.


\subsection{Production cross sections}
\label{sec:phenom5}

In order to estimate the total hadronic cross sections for the various production channels, we need to convolute the partonic cross sections with the corresponding parton distribution functions (PDFs). Here we will use the MSTW set \cite{PDFs}. Moreover, we will compute the cross sections at the NLO; {\it i.e.}, including the LO QCD corrections, for which simple analytical expressions can be obtained \cite{DJ1,DJ2}.
For the $q_u\bar{q}_d\to H^+\varphi_i^0$ associated production, the $\cO(\alpha_s)$ contributions simply correspond to the QCD corrections to the Drell-Yan process $q_u\bar{q}_d\to W^{*}$, integrating over the virtuality of the W boson.
As for the $H^+W^-$ associated production, the needed QCD corrections can be easily extracted from the SM Higgs production channels $q\bar q\to h$ and
$gg\to h$. At the LHC, $gg\to H^+W^-$ production dominates over $q\bar q\to H^+W^-$. For typical LHC hadronic center-of-mass energies, {\it i.e.}, $\sqrt{s}\sim 14$ TeV, the latter only corresponds at LO to a few percent of the total $pp\to H^+W^-$ cross section, so we can safely neglect it. The detailed expressions of the hadronic cross sections and the
QCD corrections are given in appendices~\ref{QCDcorrections} and \ref{QCDcorrections2}.
In order to estimate the theoretical uncertainty of the QCD enhancement factor
$K\equiv \sigma_{NLO}/ \sigma_{LO}$, we vary the factorization ($\mu_F$) and renormalization ($\mu_R$) scales for $\sigma_{NLO}$, keeping both scales fixed at their central value $\mu_F=\mu_R=\hat{s}$ for $\sigma_{LO}$.

When one of the intermediate scalar bosons reaches its on-shell kinematical region, one needs to estimate also its total decay rate. The explicit expressions for the tree-level scalar decay rates are presented in appendix \ref{app:decayrates}.

\subsubsection{$H^+\varphi_i^0$ associated production}

\begin{figure}[t]
\centering
\includegraphics[scale=0.56]{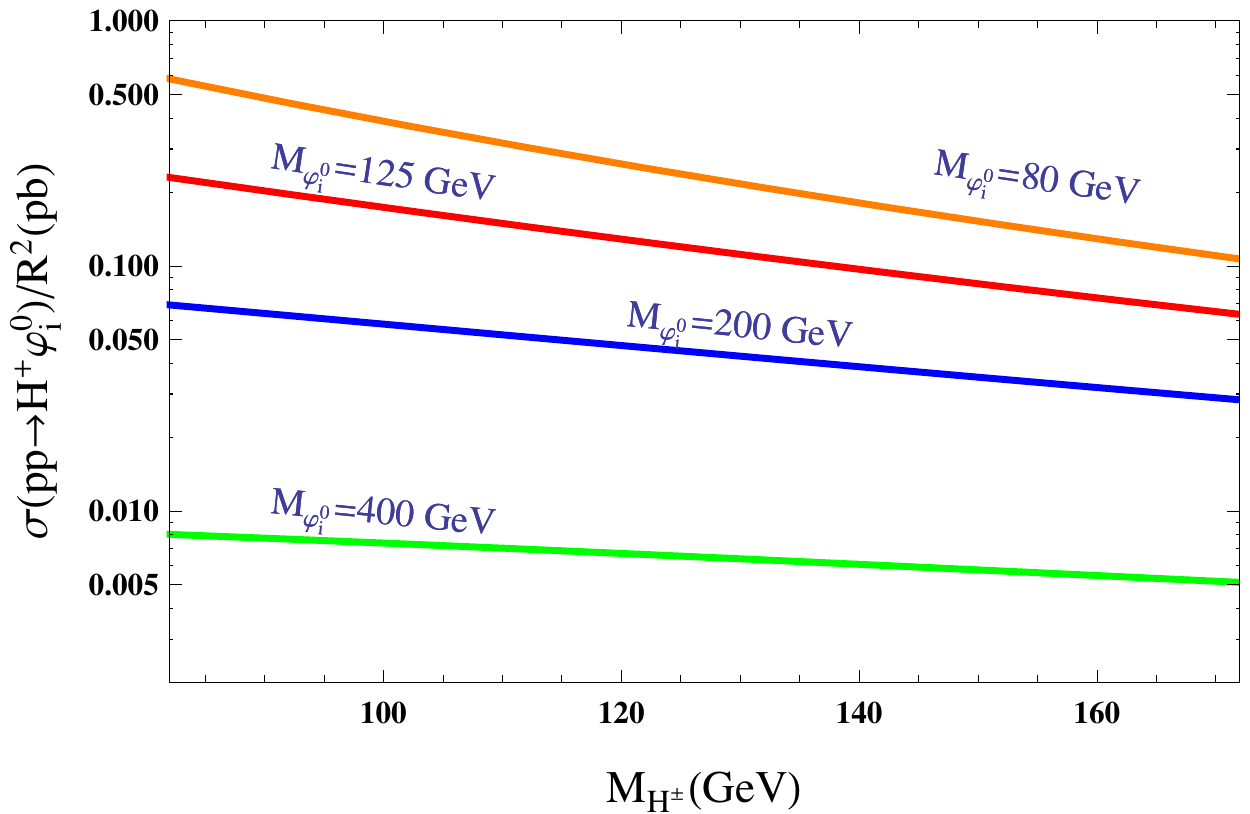}  \;\;\;  \includegraphics[scale=0.54]{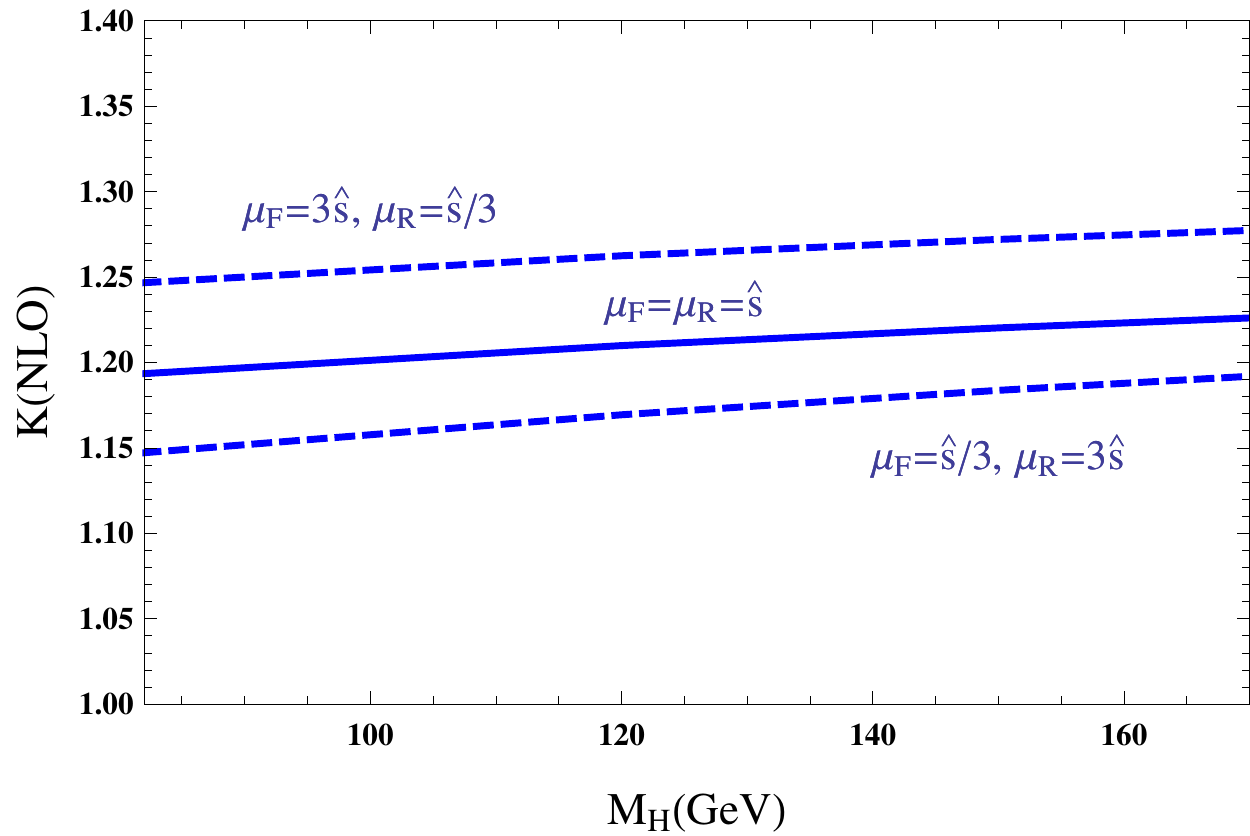}
\caption{\it LO production cross section $\sigma(pp\to H^+\varphi^0_i)/R^2$ at $\sqrt{s}=14$~TeV (left), as function of $M_{H^\pm}$, for different values of
$M_{\varphi_i^0}$. The
QCD K factor is shown (right) for $M_{\varphi_i^0}=125$ GeV and different choices of $\mu_R$ and $\mu_F$}
\label{crossDY}
\end{figure}

Assuming the most general scalar potential, the LO partonic cross section, given in Eq.~(\ref{drell-yan}), is proportional to the combination of rotation matrix elements $R^2\equiv(\mathcal{R}_{i2}^2+\mathcal{R}_{i3}^2)$. We take away the explicit dependence on the scalar-potential parameters, plotting in Fig.~\ref{crossDY} (left) the ratio
$\sigma(pp\to H^+\varphi_i^0)/R^2$ at $\sqrt{s}=14$~TeV, as a function of $M_{H^\pm}$, for different values of $M_{\varphi_i^0}$ which can be interpreted as the mass of any of the three neutral scalars of the theory.

As expected, the cross section reaches higher values for lower scalar masses. The most interesting case is of course $M_{\varphi_i^0}$=125 GeV, which could constitute a very good detection channel, since we already know that there is one scalar with that mass. If we consider $\varphi_i^0$ to be the light CP-even scalar of the theory, the cross section is suppressed by a factor $R^2=\sin^2{\tilde\alpha}$. The measurement of this production channel can be experimentally challenging due to the small value of the cross section.

QCD corrections provide a mild enhancement of the cross section. The resulting QCD K factor is shown in Fig.~\ref{crossDY} (right), for $M_{\varphi_i^0}=125$ GeV and different choices of $\mu_R$ and $\mu_F$. Its central value is around 1.2, similarly to other cross sections of the Drell-Yan type.

\subsubsection{$H^+W^-$ associated production}

For this specific production channel we are going to consider two
alternative possibilities: we can either identify the 125 GeV boson with the
lightest CP-even scalar $h$, or with the heaviest one $H$. In the first case ($M_h=125$ GeV), the scalar $H$ can be heavy enough to reach the on-shell region and, therefore, it is necessary to regulate the propagator pole with its total decay width. In the second case ($M_H=125$ GeV), both $M_h$ and $M_H$ are below the $H^+W^-$ production threshold for the whole considered range of charged Higgs masses. Therefore, there is no need to regulate the $h$ and $H$ poles (assuming their total decay widths to be small).\\

\vskip .5cm\noindent {\bf A) $\mathbf{M_h=125}$ GeV.}

Let us first estimate the size of the $H$ decay width for three representative values of $M_H$ (150, 200 and 400 GeV) and different choices for the cubic scalar couplings.
The CP-odd mass $M_A$ will always be taken within the 68\% CL range allowed by the oblique parameters. In the following discussion, we set $\cos\tilde\alpha=0.9$ and ignore the loop-induced decays $H\to gg$ and $H\to\gamma\gamma$, which are suppressed by a $\sin^2\tilde\alpha$ factor with respect to the SM.

\begin{figure}[t]
\centering
\includegraphics[scale=0.55]{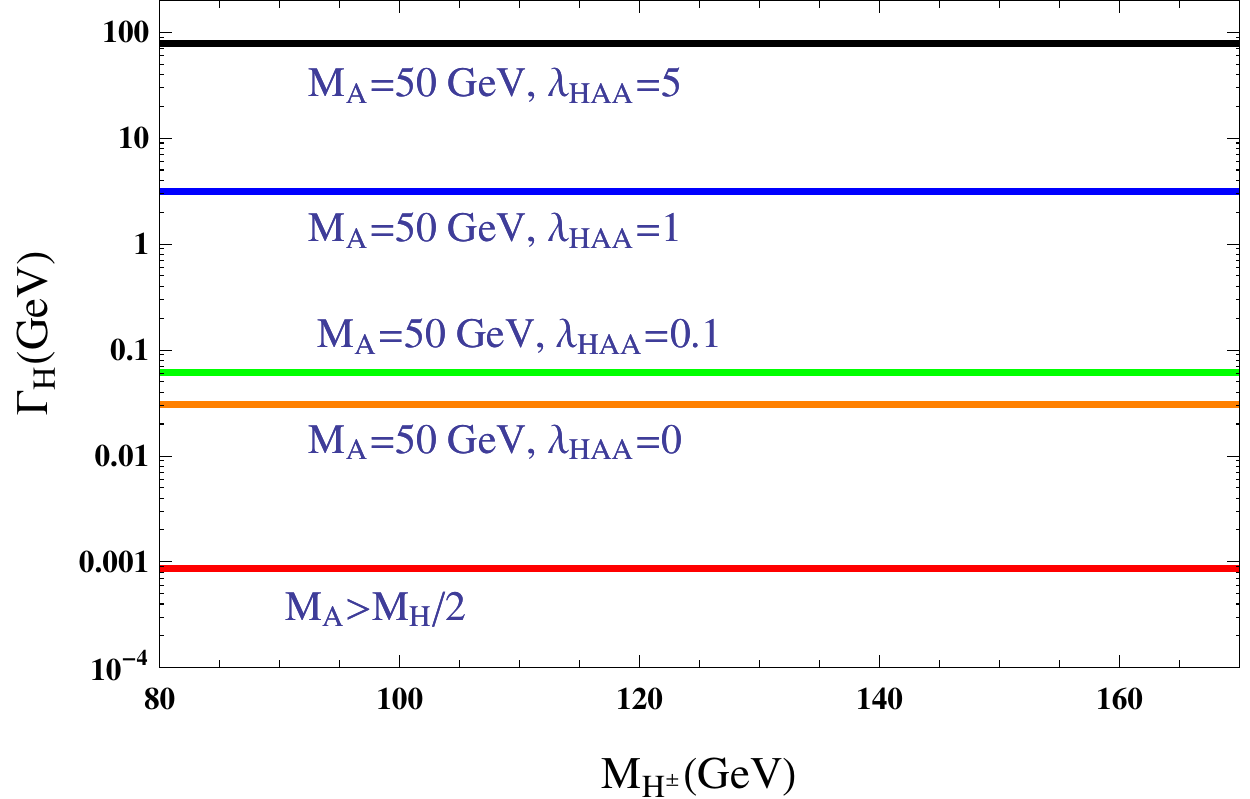} \;\;\; \includegraphics[scale=0.55]{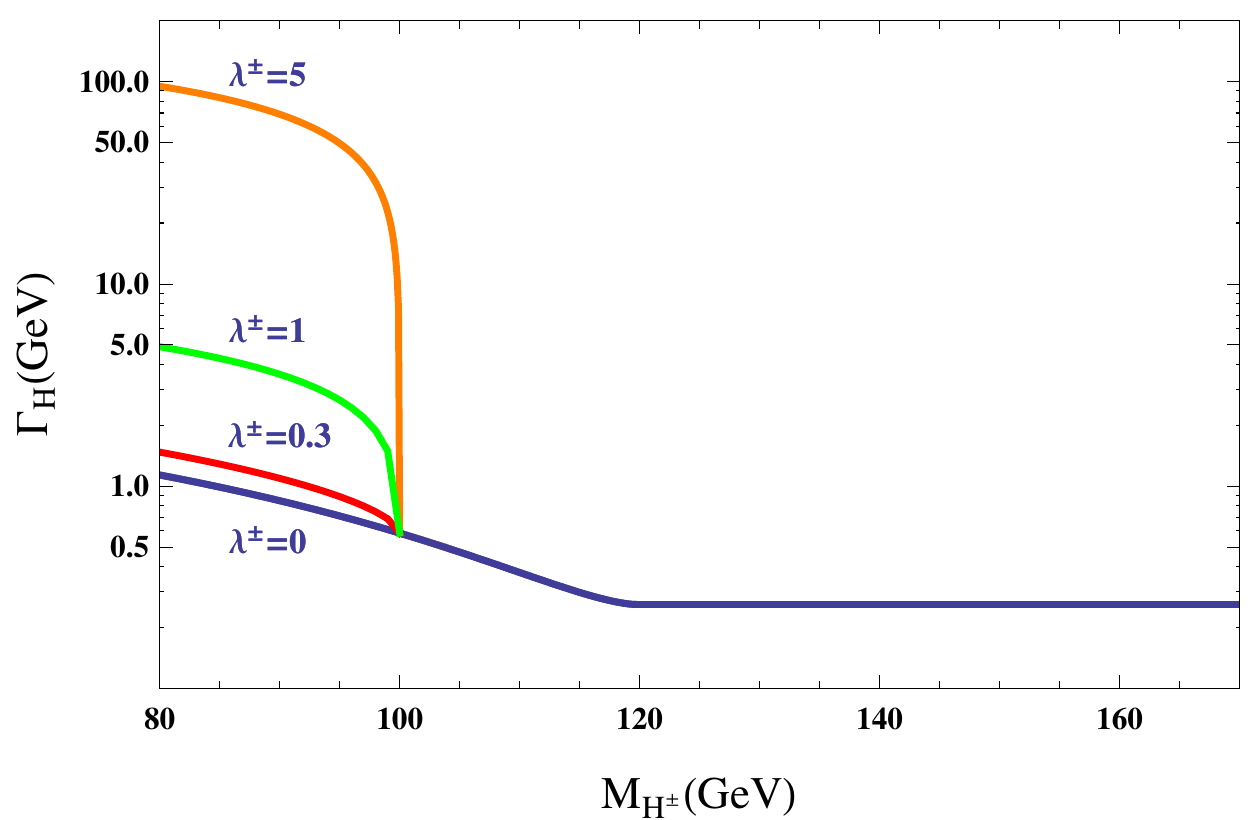} \\[2ex]
\includegraphics[scale=0.55]{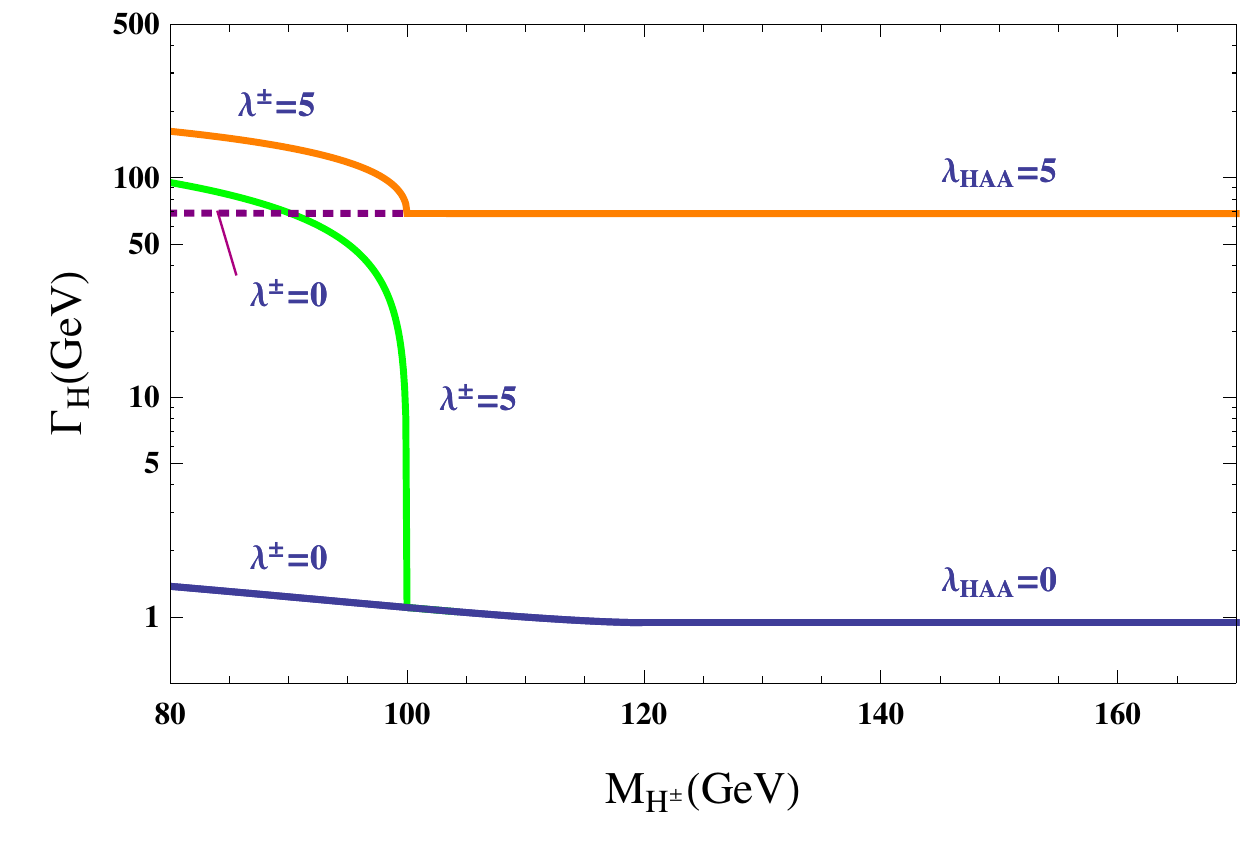} \;\;\; \includegraphics[scale=0.55]{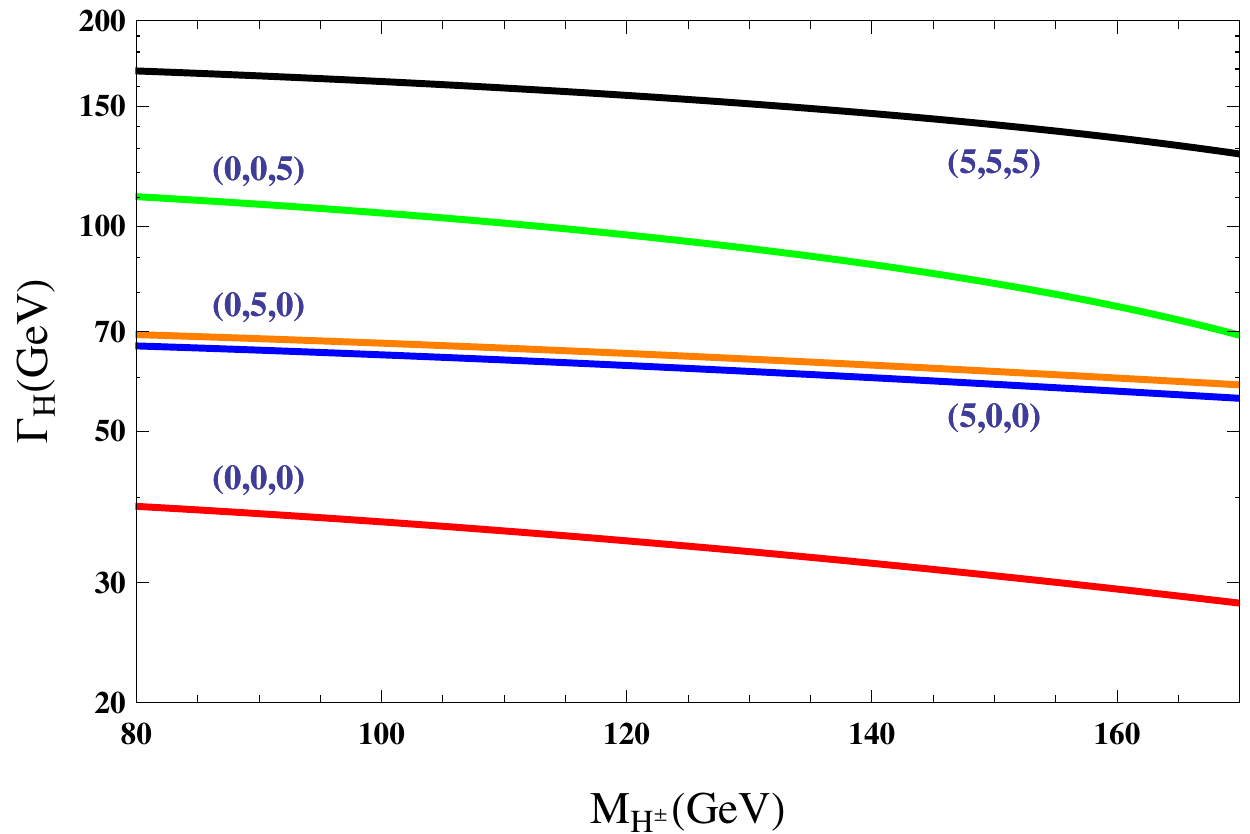}
\caption{\it Total $H$ decay rate as a function of $M_{H^{\pm}}$ for a) $M_H = 150 $ GeV with different values of $M_A$ and $|\lambda_{HAA}|$ (top-left), b) $M_H = 200 $ GeV and $M_A > M_H-M_Z$ with different values of $\lambda^{\pm}\equiv |\lambda_{HH^+H^-}|$ (top-right), c) $M_H = 200 $ GeV and $M_A = 50$ GeV with different values of $|\lambda_{HH^+H^-}|$ and $|\lambda_{HAA}|$ (bottom-left), and d) $M_H = 400 $ GeV and $M_A = 140$ GeV with different values for the set of couplings $(|\lambda_{HAA}|,\; |\lambda_{Hhh}|, \; |\lambda_{HH^+H^-}|)$ (bottom-right).}
\label{Hdecay}
\end{figure}

For $M_H=150$ GeV, the $H$ boson does not reach the on-shell region (its mass is below the $H^+W^-$ threshold) and its total decay width is in principle not needed to regulate the propagator pole. However, $\Gamma_H$ can induce sizeable effects for
small $M_A$
and large values of the cubic coupling $\lambda_{HAA}$.
This is shown in Fig.~\ref{Hdecay} (upper-left). When $M_A>M_H/2$, the $H$ width is small because its only relevant tree-level decays are $H\to b\bar b,\, WW$ and $ZZ$. However,
extra decay channels like $H\to AA$ or $H\to AZ$ are open when one allows $A$ to be light. This possibility is exemplified in the figure, taking $M_A=50$ GeV and $\lambda_{HAA}=0$ (therefore $H\to AZ$ is the only extra channel), and also for  $|\lambda_{HAA}|=0.1, \; 1$ and 5. The width $\Gamma_H$ varies roughly from around $10^{-3}$ up to 100 GeV for the considered parameter configurations.

Let us now consider $M_H=200$ GeV. If the CP-odd boson satisfies $M_A>M_H-M_Z\approx 110$ GeV, then the channels $H\to AA, \, AZ$ are closed. The open decay channels are $H\to b\bar b, \, WW, \, ZZ$ as before, plus two extra ones: $H\to H^{\pm}W^{\mp}$ (up to $M_{H^\pm} \approx 120$ GeV) and $H\to H^+H^-$ (up to $M_{H^\pm}=100$ GeV). When kinematically allowed (and if $|\lambda_{HH^+H^-}|$ is not too small), the decay into two charged scalars is the dominating channel. There is also a sizeable contribution from $H\to H^{\pm}W^{\mp}$ when this decay mode is open. The predicted values of $\Gamma_H$ are shown in Fig.~\ref{Hdecay} (upper-right) for different values of $|\lambda_{HH^+H^-}|$. If we take instead $M_A=50$ GeV, the channels $H\to AA,\, AZ$ open. The $H$ decay width is shown for this configuration in Fig.~\ref{Hdecay} (lower-left), as a function of the charged Higgs mass, taking $|\lambda_{HAA}|=0,\, 5$ and $|\lambda_{HH^+H^-}|=0,\, 5$. The total $H$ decay width obviously increases with increasing
values of $|\lambda_{HAA}|$ and $|\lambda_{HH^+H^-}|$. In the considered range of cubic couplings, $\Gamma_H$ can vary between 1 and 200 (70) GeV when $H \to H^+H^-$ is allowed (forbidden, $M_{H^\pm}>M_H/2$).

Taking a heavier mass $M_H=400$ GeV, the electroweak oblique parameters imply very stringent restrictions on $M_A$: the only value that roughly satisfies these constraints for the whole considered range of the charged Higgs mass is $M_A=140$ GeV. For this configuration, all the channels we have considered before are kinematically allowed. Besides, there is an extra one, the decay into two light CP-even scalars $H\to hh$. Thus, we have three unknown couplings $\lambda_{HAA},\; \lambda_{Hhh},$ and $\lambda_{HH^+H^-}$. The lower-right panel in Fig.~\ref{Hdecay} shows the resulting values of $\Gamma_H$, taking
$(|\lambda_{HAA}|,\; |\lambda_{Hhh}|, \;|\lambda_{HH^+H^-}|)=(0,0,0),\; (5,0,0),\; (0,5,0),\; (0,0,5),$ and $(5,5,5)$.
The total $H$ decay rate grows from around 30 GeV when the three cubic scalar couplings are zero, up to approximately 150 GeV when their values are $(5,5,5)$.

\begin{figure}[t]
\centering
\includegraphics[scale=0.55]{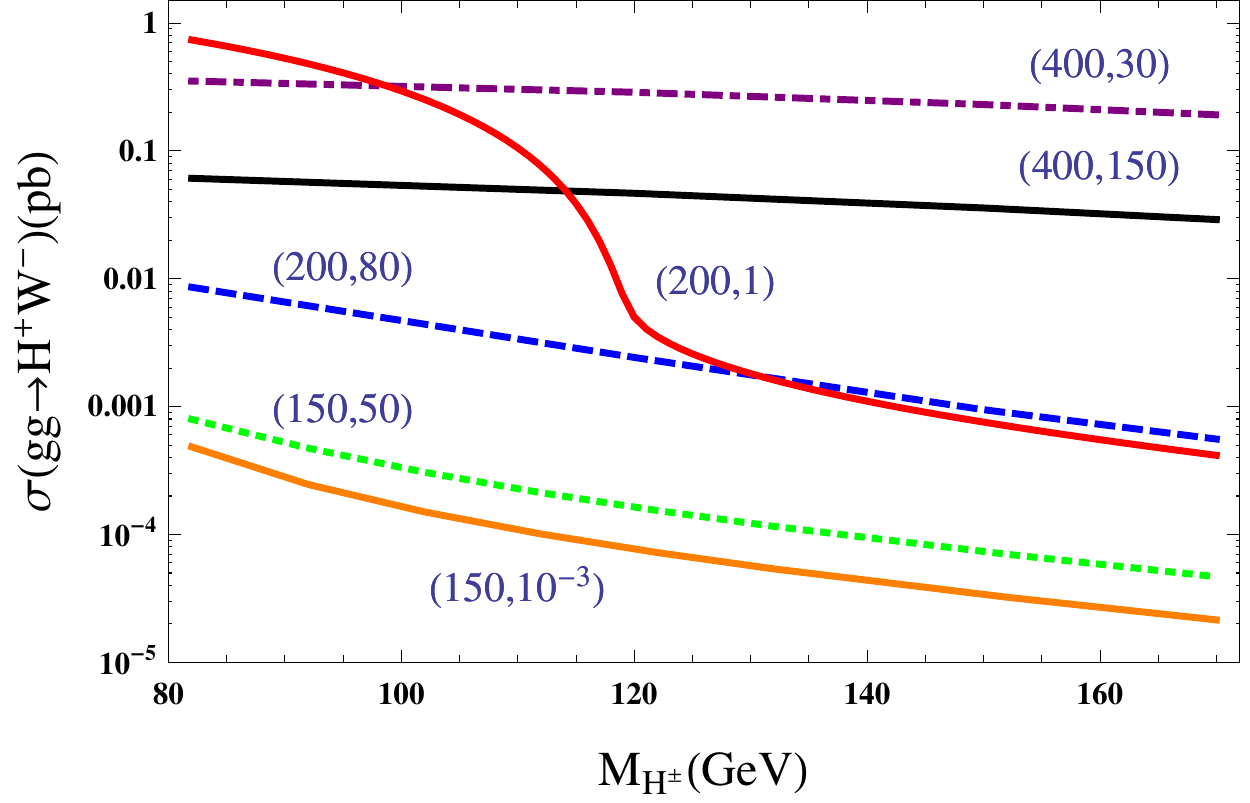} \qquad \includegraphics[scale=0.54]{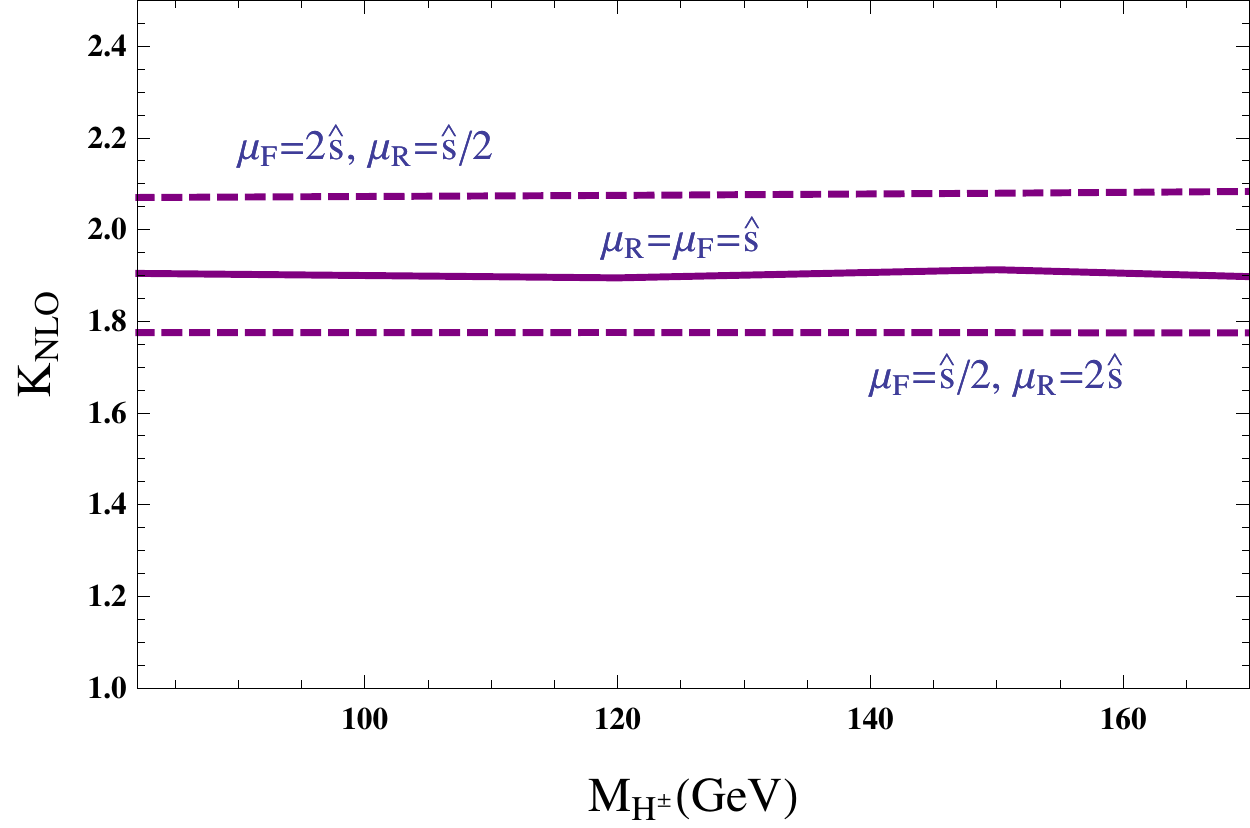}
\caption{\it LO production cross section $\sigma(pp\to H^+W^-)$ at $\sqrt{s}=14$ TeV (left), as a function of $M_{H^\pm}$, for $M_h=125$ GeV, $\cos\tilde\alpha=0.9$ and different values for the pair $(M_H,\; \Gamma_H)$ in GeV. The QCD K factor is shown (right) for $(M_H, \; \Gamma_H)=(400,30)$ GeV and different choices of $\mu_R$ and $\mu_F$.}
\label{cross1}
\end{figure}

Fig.~\ref{cross1} (left) shows the predicted LO production cross sections at $\sqrt{s}=14$ TeV,
for representative values of $M_H$ and $\Gamma_H$, which cover the range of possibilities we have just discussed:
$(M_H,\; \Gamma_H)= (150,10^{-3})$, $(150,50)$, $(200,1)$, $(200,80)$, $(400,30)$, and $(400,150)$ GeV.
The cross section is very small when both CP-even scalars are off-shell. For $M_H = 150$ GeV, $\sigma(pp\to H^+W^-)$ is roughly smaller than $10^{-3}$ pb.
With $M_H=200$ GeV and a large decay width $\Gamma_H=80$ GeV, the cross section stays below $10^{-2}$ pb; however,
with a smaller width $\Gamma_H=1$ GeV, the cross section is enhanced by approximately two orders of magnitude
(three orders of magnitude with respect to the previous cases), in the region where $M_H$ is on-shell ($M_{H^\pm} \lesssim 120$ GeV).

The most interesting case is when $M_H=400$ GeV, because the cross section gets enhanced by the on-shell $H$ pole, reaching higher values around 0.1~pb.
The QCD K factor for this $H$ mass and $\Gamma_H=30$~GeV is given in Fig.~\ref{cross1} (right), and it is practically constant in the whole range of $M_{H^\pm}$; it approximately corresponds to the K factor for the production of a SM Higgs with a 400 GeV mass. Its central value is around 1.9. A very similar K factor is obtained for $\Gamma_H=150$~GeV, although with a smaller cross section.

Thus, a heavy $H$ boson would be the most favourable situation from the experimental point of view, with production cross sections between $10^{-2}$ and 1 pb at $\sqrt{s}= 14$~TeV, depending on the value of $\Gamma_H$, which are potentially measurable at the LHC. As we have seen, they are increased by a factor of $\approx$ 2 by the NLO QCD corrections. For the other configurations both CP-even scalars are off-shell and the value of the cross section decreases by a few orders of magnitude, which results pretty challenging for the LHC, if not impossible. Nonetheless, these small values could turn out to be measurable in the future if the LHC luminosity is increased by a factor of 10, as planned for its High-Luminosity option.

\vskip .5cm\noindent {\bf B) $\mathbf{M_H=125}$ GeV.}

\begin{figure}[t]
\centering
\includegraphics[scale=0.54]{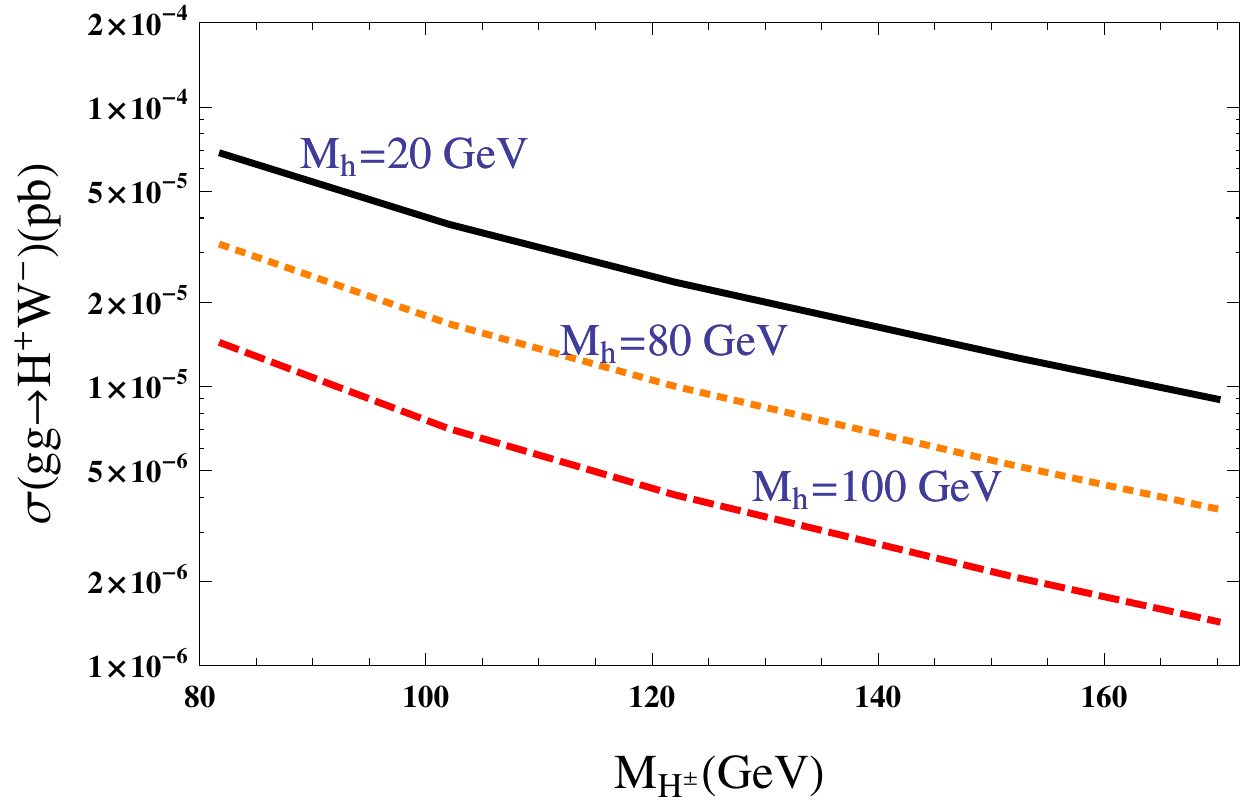} \qquad  \includegraphics[scale=0.52]{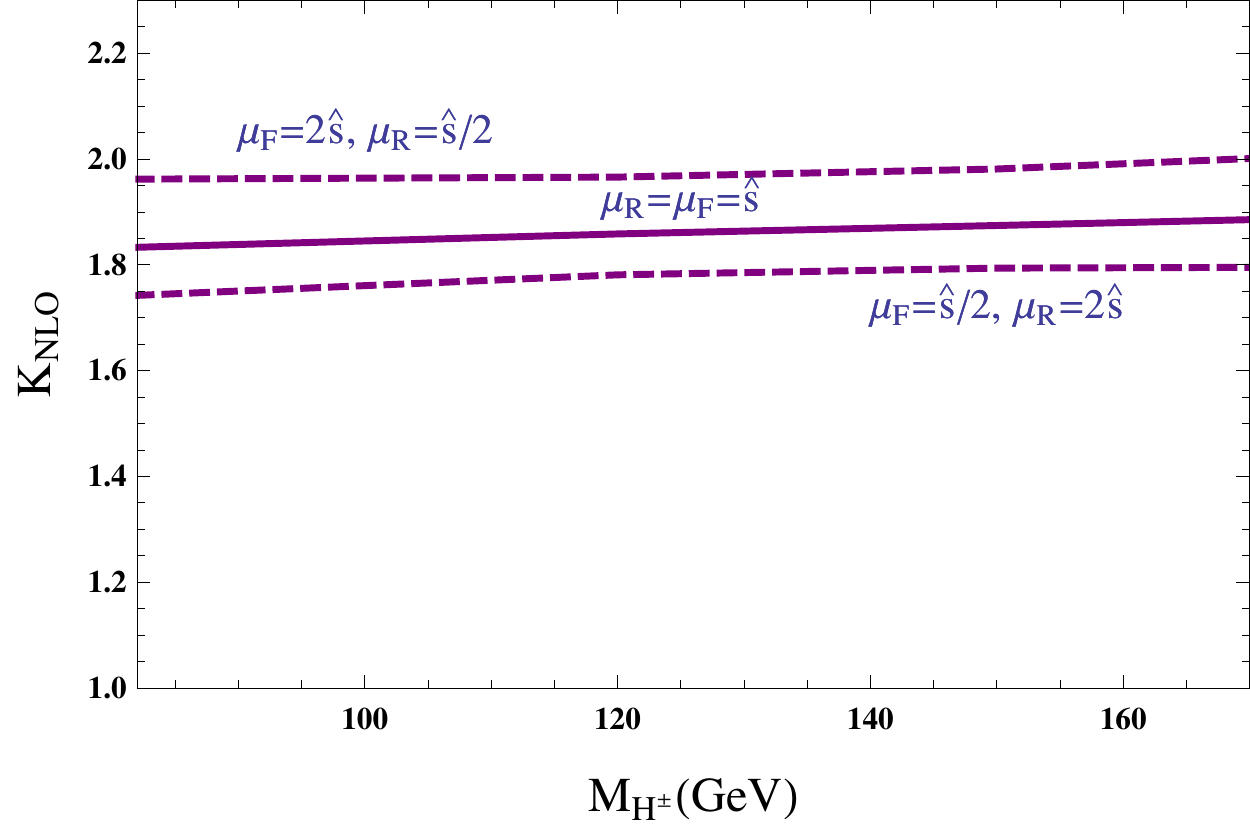}
\caption{\it LO production cross section $\sigma(pp\to H^+W^-)$ at $\sqrt{s}=14$ TeV (left), as a function of $M_{H^\pm}$, for $M_H=125$ GeV, $\sin\tilde\alpha=0.99$ and $M_h=20$, $80$, $100$ GeV. The NLO QCD K factor (right) is shown for $M_h=20$ GeV and different choices of $\mu_R$ and $\mu_F$.}
\label{cross2}
\end{figure}

In this case both CP-even neutral scalars are off-shell and their decay widths can be neglected (assuming they are small). The scalar mixing angle must be small enough to avoid the LEP constraints, thus we take $\sin\tilde\alpha=0.99$, as we have done before in the analysis of branching ratios. The mass of the light scalar will be set to $M_h=20$, $80$ and $100$ GeV. The predicted LO production cross sections at $\sqrt{s}=14$~TeV are shown in Fig.~\ref{cross2} (left). For the chosen values of $M_h$, they range in between $10^{-5}$ and $10^{-6}$ pb. These values are extremely small and lay below the experimental sensitivity attainable in the near future. This scenario is thus, the most challenging experimentally. The computed K factor, Fig~\ref{cross2} (right), has
a similar value to the one obtained in the previous scenario.


\section{Conclusions}

The recent discovery of a Higgs-like boson has confirmed the existence of a scalar sector, which so far seems compatible with the SM predictions.
As it is widely known, an enlarged scalar sector is not forbidden by the symmetries of the electroweak theory, and there exists a broad range of possibilities
satisfying all experimental constraints. The direct discovery of another scalar particle would represent a major break-through in
particle physics, opening a window into a new high-energy dynamics and providing priceless information on which type of extension,
amongst many theoretical models of the scalar sector, is preferred by Nature.

Here we have focused on a particular 2HDM scenario, characterized by a fermiophobic charged Higgs, which would have evaded all experimental searches
performed until now. It is a quite predictive case, since all Yukawa couplings are determined by the mixing among the neutral scalars.
We have assumed a CP-conserving scalar potential and have restricted our analysis to the range $M_{H^\pm}\in\left[ M_W, M_W+M_Z\right]$, so that
only a few decay modes are kinematically open. We have presented detailed formulae for the loop-induced decay $H^+ \to W^+\gamma$, which becomes very relevant
in this mass region, and for the tree-level three-body decays of the charged scalar. We have analyzed the parameter space of the model, in order to
characterize the possible values of the $H^\pm$ decay width and branching ratios, taking into account the constraints from LHC, LEP and flavour data.

The two most important production channels for a fermiophobic charged scalar have been investigated, including NLO QCD corrections:
the associated production with either a neutral scalar or a charged $W$; {\it i.e.}, $q_u\bar q_d\to H^+\varphi_i^0$ and
$gg\to H^+W^-$. The predicted cross sections are small in most of the parameter space, making the experimental search challenging,
but they become very sizeable ($\ge 10^{-3}$~pb) for large values of the mass of the heavy neutral scalar $H$. In some extreme cases, cross sections between
0.1 and 1 pb are obtained. Thus, the detection of a fermiophobic $H^\pm$ at the LHC seems plausible in the near future. The interesting features of this possible scenario should encourage specific experimental searches for such a particle in the LHC data.


\begin{appendix}

\section{Scalar Potential}
\label{app:potential}

In the Higgs basis, the most general scalar potential takes the form
\beqn\label{eq:potential}
V & = & \mu_1\; \Phi_1^\dagger\Phi_1\, +\, \mu_2\; \Phi_2^\dagger\Phi_2 \, +\, \left[\mu_3\; \Phi_1^\dagger\Phi_2 \, +\, \mu_3^*\; \Phi_2^\dagger\Phi_1\right]
\no\\ & + & \lambda_1\, \left(\Phi_1^\dagger\Phi_1\right)^2 \, +\, \lambda_2\, \left(\Phi_2^\dagger\Phi_2\right)^2 \, +\,
\lambda_3\, \left(\Phi_1^\dagger\Phi_1\right) \left(\Phi_2^\dagger\Phi_2\right) \, +\, \lambda_4\, \left(\Phi_1^\dagger\Phi_2\right) \left(\Phi_2^\dagger\Phi_1\right)
\no\\ & + & \left[  \left(\lambda_5\; \Phi_1^\dagger\Phi_2 \, +\,\lambda_6\; \Phi_1^\dagger\Phi_1 \, +\,\lambda_7\; \Phi_2^\dagger\Phi_2\right) \left(\Phi_1^\dagger\Phi_2\right)
\, +\, \mathrm{h.c.}\right]\, .
\eeqn
The hermiticity of the potential requires all parameters to be real except $\mu_3$, $\lambda_5$, $\lambda_6$ and $\lambda_7$; thus, there are 14 real parameters.
The minimization conditions
$\langle 0|\Phi_1^T(x)|0\rangle =\frac{1}{\sqrt{2}}\, (0, v)$ and $\langle 0|\Phi_2^T(x)|0\rangle =\frac{1}{\sqrt{2}}\, (0, 0)$
impose the relations
\bel{eq:minimum}
\mu_1\; =\; -\lambda_1\, v^2\, ,
\qquad\qquad\qquad
\mu_3\; =\; -\frac{1}{2}\,\lambda_6\, v^2\, .
\ee
The potential can then be decomposed into a quadratic term plus cubic and quartic interactions
\bel{eq:potential2}
V\; =\; -\frac{1}{4}\,\lambda_1\, v^4\, +\, V_2 \, +\,  V_3 \, +\,  V_4\, .
\ee
The mass terms take the form
\beqn\label{eq:mass_term}
V_2 & = &
M_{H^\pm}^2\, H^+ H^-\, +\, \frac{1}{2}\, \left(S_1, S_2, S_3\right)\; \mathcal{M}\; \left(\ba S_1\\ S_2\\ S_3\ea\right)
\no\\[10pt] & = &
M_{H^\pm}^2\, H^+ H^-\, +\, \frac{1}{2}\, M_h^2\, h^2\, +\, \frac{1}{2}\, M_H^2\, H^2\, +\, \frac{1}{2}\, M_A^2\, A^2\, ,
\eeqn
with
\bel{eq:mplus}
M_{H^\pm}^2\; =\; \mu_2 + \frac{1}{2}\,\lambda_3\, v^2
\ee
and
\bel{eq:mass_matrix}
\mathcal{M}\; =\; \left(\begin{array}{ccc}
2\lambda_1 v^2 & v^2\, \lambda_6^{\mathrm{R}} & -v^2\, \lambda_6^{\mathrm{I}}\\
v^2\, \lambda_6^{\mathrm{R}} & M_{H^\pm}^2  + v^2\left(\frac{\lambda_4}{2} + \lambda_5^{\mathrm{R}}\right)
& -v^2\, \lambda_5^{\mathrm{I}}\\
-v^2\, \lambda_6^{\mathrm{I}} & -v^2\, \lambda_5^{\mathrm{I}} & M_{H^\pm}^2  +  v^2\left(\frac{\lambda_4}{2} - \lambda_5^{\mathrm{R}}\right)
\ea\right)\, ,
\ee
where $\lambda_i^{\mathrm{R}}\equiv \mathrm{Re}(\lambda_i)$ and $\lambda_i^{\mathrm{I}}\equiv \mathrm{Im}(\lambda_i)$.
The symmetric mass matrix $\mathcal{M}$ is diagonalized by an orthogonal matrix $\mathcal{R}$, which defines the neutral mass eigenstates:
\bel{eq:mass_diagonalization}
\mathcal{M}\; =\; \mathcal{R}^T\; \mathcal{M}_D \; \mathcal{R}\, ,
\qquad\qquad\qquad
\varphi^0 \; =\; \mathcal{R}\; S \, ,
\ee
where we have introduced the shorthand matrix notation
\bel{eq:mass_diagonalization}
\mathcal{M}_D \; \equiv \; \left(\begin{array}{ccc} M_h^2 & 0 & 0 \\ 0 & M_H^2 &  0 \\ 0 & 0 & M_A^2
\ea\right)\; \, ,
\qquad\qquad
\varphi^0 \; \equiv \; \left(\ba h\\ H\\ A\ea\right)\;
\; \, ,
\qquad\qquad S \; \equiv \; \left(\ba S_1\\ S_2\\ S_3\ea\right)\, .
\ee
Since the trace remains invariant, the masses satisfy the relation
\bel{eq:mass_sum}
M_h^2 \, +\, M_H^2 \, +\, M_A^2\; =\; 2\, M_{H^\pm}^2\, +\, v^2\,\left(2\,\lambda_1 +\lambda_4\right)\, .
\ee

The minimization conditions allow us to trade the parameters $\mu_1$ and $\mu_3$ by $v$ and $\lambda_6$. The freedom to rephase the field $\Phi_2$ implies,
moreover, that only the relative phases among $\lambda_5$, $\lambda_6$ and $\lambda_7$ are physical; but only two of them are independent. Therefore,
we can fully characterize the potential with 11 parameters: $v$, $\mu_2$, $|\lambda_{1,\ldots,7}|$,
$\mathrm{arg}(\lambda_5\lambda_6^*)$ and $\mathrm{arg}(\lambda_5\lambda_7^*)$. Four parameters can be determined through the physical scalar masses \cite{ilisie1}. The matrix equation
\begin{align}
(\mathcal{M} \; \mathcal{R}^T - \mathcal{R}^T \; \mathcal{M}_D)  \; 
= \; 0
\end{align}
relates the scalar masses and mixings. Summing the second row with $(-i)$ times the third row, one obtains the identity (imaginary parts included):
\begin{align}
v^2 \lambda_6 \mathcal{R}_{i1} + \Big[ M_{H^\pm}^2 - M_{\varphi_i^0}^2 + v^2 \Big( \frac{\lambda_4}{2} + \lambda_5 \Big)  \Big]\,
(\mathcal{R}_{i2} - i \mathcal{R}_{i3}) + 2 iv^2\lambda_5 \mathcal{R}_{i3}\; =\; 0\, .
\label{2+3rows}
\end{align}
This proves in full generality that
\begin{align}
(\mathcal{R}_{i2}-i\mathcal{R}_{i3})\; \frac{M_{\varphi_i^0}^2-M_{H^\pm}^2}{v^2}\; =\; (\mathcal{R}_{i2}-i\mathcal{R}_{i3})\,
\Big( \frac{\lambda_4}{2} + \lambda_5 \Big) + 2 i \mathcal{R}_{i3} \lambda_5 + \mathcal{R}_{i1} \lambda_6
\; =\; \lambda_{H^+ G^-\varphi_i^0}\, .
\end{align}
Taking instead the first row, one gets:
\begin{align}
\big( 2 \lambda_1  v^2 - M_{\varphi_i^0}^2\big) \; \mathcal{R}_{i1} + v^2 \lambda_6^{\text{R}} \mathcal{R}_{i2} - v^2 \lambda_6^{\text{I}} \mathcal{R}_{i3}\; =\; 0\, ,
\label{firstrow}
\end{align}
which generalizes the usual relation determining $\tan\tilde\alpha$ in the CP-conserving limit ($\mathcal{R}_{13}=\mathcal{R}_{23}=0$). It also proves that the following identity holds in general
\begin{align}
\frac{M_{\varphi_i^0}^2}{v^2} \; \mathcal{R}_{i1} \; = \; 2 R_{i1} \lambda_1 + i \mathcal{R}_{i3} \lambda_6 + (\mathcal{R}_{i2}-i \mathcal{R}_{i3})\lambda_6^{\text{R}}\; =\; \lambda_{G^+G^-\varphi_i^0}\, .
\end{align}
Here, similarly to Eq.~\eqn{hHPHM}, we have parametrized the Goldstone terms of $V_3$ in the form
\begin{align}
\Big(\; v \; \lambda_{H^+ G^-\varphi_i^0} \; H^+ G^- \varphi_i^0 \;   + \text{h.c.} \; \Big) \; +
 \; v \; \lambda_{G^+ G^-\varphi_i^0}  \; G^+ G^- \varphi_i^0 \;     \; \subset  \; V_3\, .
\end{align}
These identities generalize the ones from \cite{HaberCP}, that are valid only in the CP-conserving limit of the scalar potential. They turn out to be very useful if one works in $R_\xi$ gauges with a fully general potential.

Using again Eq.~(\ref{firstrow}), the orthogonality of $\mathcal{R}$ implies:
\begin{align}
\sum_i \mathcal{R}_{i1}^2\; M_{\varphi_i^0}^2\; =\; 2\lambda_1 v^2 \, ,
\qquad\;
\sum_i \mathcal{R}_{i1} \mathcal{R}_{i2} \; M_{\varphi_i^0}^2\; =\; \lambda_6^{\text{R}} v^2 \, ,
\qquad\;
\sum_i \mathcal{R}_{i1}\mathcal{R}_{i3} \; M_{\varphi_i^0}^2\; =\; - \lambda_6^{\text{I}} v^2 \, .
\label{ortogonal}
\end{align}
Eq.~(\ref{2+3rows}) gives the additional orthogonality relations.
\begin{align}
\sum_i \mathcal{R}_{i1} (\mathcal{R}_{i2}-i\mathcal{R}_{i3})\; M_{\varphi_i^0}^2 \; &= \; \lambda_6 v^2\, ,
\\
\sum_i \mathcal{R}_{i2} (\mathcal{R}_{i2}-i\mathcal{R}_{i3}) \; M_{\varphi_i^0}^2 \; &= \; M_{H^\pm}^2 +  v^2 \Big( \frac{\lambda_4}{2} + \lambda_5 \Big)\, ,
\\
i\sum_i \mathcal{R}_{i3} (\mathcal{R}_{i2}-i\mathcal{R}_{i3}) \; M_{\varphi_i^0}^2 \; &= \;  M_{H^\pm}^2 +  v^2 \Big( \frac{\lambda_4}{2} - \lambda_5 \Big)\, .
\end{align}
The first identity reproduces in complex form the last two real equations in (\ref{ortogonal}). Separating the real and imaginary parts of the last two relations, one gets:
\begin{align}
& \sum_i \mathcal{R}_{i2}^2 \; M_{\varphi_i^0}^2 \; = \; M_{H^\pm}^2 + v^2 \Big( \frac{\lambda_4}{2} + \lambda_5^{\text{R}} \Big) \; , \\
& \sum_i \mathcal{R}_{i3}^2 \; M_{\varphi_i^0}^2 \; = \; M_{H^\pm}^2 + v^2 \Big( \frac{\lambda_4}{2} - \lambda_5^{\text{R}} \Big) \; , \\
& \sum_i \mathcal{R}_{i2} \mathcal{R}_{i3} \; M_{\varphi_i^0}^2 \; = - v^2  \lambda_5^{\text{I}} \; .
\end{align}

\subsection{Inert 2HDM}
\label{subsec:inert}

Imposing a discrete ${\cal Z}_2$ symmetry such that all SM fields remain invariant under a ${\cal Z}_2$ transformation, while
\bel{eq:Z2sym}
\Phi_1\;\to\; \Phi_1\, ,
\qquad\qquad
\Phi_2\;\to\; -\Phi_2\, ,
\ee
one makes the second scalar doublet {\it inert}\/: linear interactions of $\Phi_2$ with the SM fields are odd under a ${\cal Z}_2$ transformation, and thus forbidden~\cite{Ma:2008uza,Ma:2006km}. In particular, $\Phi_2$ is fermiophobic. This inert scalar doublet can only interact with the other fields through quadratic couplings. The lightest neutral component of $\Phi_2$ is then a very good candidate for dark matter.

The ${\cal Z}_2$ symmetry implies a significant simplification of the scalar potential, because all terms with an odd number of $\Phi_2$ fields vanish:
$\mu_3=\lambda_6=\lambda_7=0$. Moreover, making an appropriate rephasing of $\Phi_2$, $\lambda_5$ can be taken real. Therefore, the neutral mass matrix \eqn{eq:mass_matrix} becomes diagonal and there is no mixing among the neutral scalars ($\cR = I$). The neutral scalar masses are given by:
\bel{eq:InertMasses}
M^2_h\; =\; 2 \lambda_1 v^2\, ,
\qquad
M^2_H\; =\; M^2_{H^\pm} + \left(\frac{\lambda_4}{2}+\lambda_5\right)\, v^2\, ,
\qquad
M^2_A\; =\; M^2_{H^\pm} + \left(\frac{\lambda_4}{2}-\lambda_5\right)\, v^2\, .
\ee
%


\section{Heavy neutral Higgs decay rates}
\label{app:decayrates}

In this section we are going to write down the tree-level on-shell two-body dominant decay rates of a heavy neutral Higgs. All the formulae presented here are, as in section \ref{sec:calc}, completely general (no assumptions are made on the Higgs potential and the A2HDM Yukawa structure is assumed). The decay rate of a neutral scalar to a pair of massive fermions is given by:
\begin{align}
 \Gamma(\varphi_i^0 \to f \bar{f})\; =\;  \frac{N_c^f \, m_f^2 \, M_{\varphi_i^0}}{8\,\pi\, v^2} \; \Big( 1-
 \frac{4m_f^2}{M_{\varphi_i^0}^2}  \Big)^{3/2} \; \Big[ \;    \text{Re}\big( y_f^{\varphi_i^0} \big)^2 + \text{Im}\big( y_f^{\varphi_i^0} \big)^2 \; \Big(1-\frac{4m_f^2}{M_{\varphi_i^0}^2} \Big)^{-1}              \; \Big]\, ,
\end{align}
where $N_c^f$ is 1 for leptons and 3 for quarks. The decay into two gauge bosons reads ($V=W,Z$)
\begin{align}
\Gamma(\varphi_i^0 \to VV)\; =\; \mathcal{R}_{i1}^2 \;\,
\frac{M_{\varphi_i^0}^3 \; \delta_V}{32 \, \pi \, v^2 }
\; \Big( 1-\frac{4M_V^2}{M_{\varphi_i^0}^2} \Big)^{1/2}
\Big( 1 - \frac{4 M_V^2}{M_{\varphi_i^0}^2} + \frac{12 M_V^4}{M_{\varphi_i^0}^4} \Big)\, ,
\end{align}
with $\delta_Z=1$ and $\delta_W=2$. Other channels that can bring important contributions are $\varphi_i^0 \to \varphi_j^0 \varphi_j^0$ and $\varphi_i^0 \to H^+H^-$. The corresponding decay widths are given by
\begin{align}
\Gamma(\varphi_i^0 \to \varphi_j^0 \varphi_j^0) &\; =\; \frac{v^2\; \lambda^2_{\varphi_i^0\varphi_j^0\varphi_j^0}}{32 \, \pi \, M_{\varphi_i^0}}\;
\Big( 1- \frac{4M_{\varphi_j^0}^2}{M_{\varphi_i^0}^2}   \Big)^{1/2}\, ,  \\
\Gamma(\varphi_i^0 \to H^+H^-) &\; =\; \frac{v^2\; \lambda^2_{\varphi_i^0 H^+ H^-}}{16 \, \pi \, M_{\varphi_i^0}}\;
\Big( 1- \frac{4M_{H^\pm}^2}{M_{\varphi_i^0}^2}   \Big)^{1/2}\, ,
\end{align}
where, for the charged Higgs interaction Lagrangian we have used the parametrization given in (\ref{hHPHM}) and we have parametrized the cubic interaction of the neutral Higgs fields as
\be
\cL_{\varphi_i^0 \varphi_j^0 \varphi_j^0 }\; =\; - \frac{v}{2} \; \lambda_{\varphi^0_i \varphi^0_j \varphi^0_j }\;\, \varphi^0_i\, \varphi^0_j\,\varphi^0_j\,\, .
\label{hHPHM2}
\ee
Explicit expressions for these couplings can be found in \cite{ilisie1}. Here we didn't consider the off-shell $\varphi_i^0 \to \varphi_j^{0*} \varphi_j^{0*}$ decay mode because in addition to its kinematical suppression it also depends on the unknown parameter
$\lambda_{\varphi^0_i \varphi^0_j \varphi^0_j }$ and would not bring useful information. The last two processes that must be taken into account are
$\varphi_i^0\to \varphi_j^0 Z$ and $\varphi_i^0 \to H^+ W^-$. We have
\begin{align}
\Gamma(\varphi_i^0\to \varphi_j^0 Z) &\; =\;
\left( \mathcal{R}_{i3}\,\mathcal{R}_{j2}-\mathcal{R}_{i2}\,\mathcal{R}_{j3}\right)^2
\; \frac{1}{16\,\pi\, v^2 M_{\varphi_i^0}^3}\;
\lambda^{3/2}( M_{\varphi_i^0}^2, M_{\varphi_j^0}^2,M_Z^2)\, , \\
\Gamma(\varphi_i^0\to H^+ W^-) &\; =\; (\mathcal{R}_{i2}^2 +  \mathcal{R}_{i3}^2) \;
\frac{1}{16\,\pi\, v^2 M_{\varphi_i^0}^3}\;
\lambda^{3/2}( M_{\varphi_i^0}^2, M_{H^\pm}^2,M_W^2) \, .
\end{align}
Again, the scalar couplings to gauge bosons are taken from \cite{ilisie1}.


\section{QCD corrections to $\mathbf{pp\to H^+\boldsymbol{\varphi}_i^0}$}
\label{QCDcorrections}

For the $H^+\varphi_i^0$ associated production, we write the LO hadronic cross section as
\be
\sigma_{\text{LO}} \; =\; \int_{\tau_0}^1 d\tau \int_\tau^1 \frac{dx}{x} \; \sum_{q_u,\bar{q}_d} \; \Big[  \;
q_u(x,\mu_F) \, \bar{q}_d(\tau/x,\mu_F)  +  \bar{q}_d(x,\mu_F) \, q_u(\tau/x,\mu_F)
 \; \Big] \; \hat{\sigma}_{\text{LO}}(\hat{s}=\tau s)\, ,
\label{crossPDFs}
\ee
where we have introduced the shorthand notation $\hat{\sigma}_{\text{LO}} \equiv \hat{\sigma}(q_u\bar{q}_d\to H^+\varphi_i^0)$, for the partonic cross section given in Eq.~(\ref{drell-yan}). As usual, the partonic invariant-mass $\hat{s}$ must be expressed as a fraction of the hadronic center-of-mass energy $s$, that is $\hat{s}=\tau s$. The lower integration limit is given by $\tau_0=(M_{H^\pm}+M_{\varphi_i^0})^2/s$. The PDFs $q_i(x,\mu_F)$, for a given quark flavour `i', depend on the momentum fraction $x$ and the factorization scale $\mu_F$.

The NLO cross section, that includes first-order QCD corrections, can be cast in the simple form \cite{DJ2,DJ1}
\be
\sigma_{\text{NLO}}\; =\; \sigma_{\text{LO}} + \Delta\sigma_{q\bar{q}} + \Delta\sigma_{qg}\, ,
\ee
where $\Delta\sigma_{q\bar{q}}$ and $\Delta\sigma_{qg}$ are given by
\begin{align}
 \Delta\sigma_{q\bar{q}} &\; =\; \frac{\alpha_s(\mu_R)}{\pi}\,  \int_{\tau_0}^1 d\tau \int_\tau^1 \frac{dx}{x}\;\sum_{q_u,\bar{q}_d}
 \; \Big[ \,
q_u(x,\mu_F) \, \bar{q}_d(\tau/x,\mu_F)  +  \bar{q}_d(x,\mu_F) \, q_u(\tau/x,\mu_F) \,
 \Big]
\notag \\
 & \qquad \times \; \int_{\tau_0/\tau}^1 \;  dz \; \hat{\sigma}_{\text{LO}} (\tau s z) \; \omega_{q\bar q}(z)
 \, , \\[2ex]
  \Delta\sigma_{qg} &\; =\; \frac{\alpha_s(\mu_R)}{\pi}\,  \int_{\tau_0}^1 d\tau \int_\tau^1 \frac{dx}{x}\;\sum_{q_u,\bar{q}_d}
  \; \Big[ \,
q_u(x,\mu_F) \, g(\tau/x,\mu_F)  +  g(x,\mu_F) \, q_u(\tau/x,\mu_F)
\notag \\
 & \qquad + \; \bar{q}_d(x,\mu_F) \, g(\tau/x,\mu_F)  +  g(x,\mu_F) \, \bar q_d(\tau/x,\mu_F) \,  \Big]  \; \int_{\tau_0/\tau}^1 \;  dz \; \hat{\sigma}_{\text{LO}} (\tau s z) \; \omega_{q g}(z) \, ,
\end{align}
with $\mu_R$ the renormalization scale and
\begin{align}
\omega_{q\bar{q}}(z) \; &= \; -P_{qq}(z)\;\log\Big(\frac{\mu_F^2}{\tau s}\Big)\,  +\, \frac{4}{3}\,\Big[
\Big( \frac{\pi^2}{3} -4 \Big)\,\delta(1-z) + 2\, (1+z^2)\,\Big( \frac{\log(1-z)}{1-z}  \Big)_{+}
\Big] \, , \notag \\[1.5ex]
\omega_{qg}(z) \; &= \; -\frac{1}{2}\, P_{qg}(z)\,\log\Big(\frac{\mu_F^2}{(1-z)^2\tau s}\Big)\, +\, \frac{1}{8}\,\Big[
1 + 6 z - 7 z^2
\Big]\, .
\end{align}
The Altarelli-Parisi splitting functions $P_{qq}$ and $P_{qg}$ are given by
\begin{align}
P_{qq}(z)\; =\; \frac{4}{3}\,\Big[ \frac{1+z^2}{(1-z)_{+}} + \frac{3}{2}\,\delta(1-z) \Big]\, , &&&
P_{qg}(z)\; =\; \frac{1}{2}\,\Big[z^2 + (1-z)^2 \Big] \, ,
\end{align}
where $F_+$ is the `$+$' distribution defined as $F_+(x)\, =\, F(x)-\delta(1-x)\int_0^1 dx'\, F(x')$, and
\be
\int_a^1 dz \; g(z) \; \Big( \frac{f(z)}{1-z} \Big)_+ \;\equiv\; \int_a^1 dz \; \Big( g(z)-g(1) \Big)\;\frac{f(z)}{1-z}\, -\, g(1)\int_0^a dz\; \frac{f(z)}{1-z}\, .
\ee
%


\section{QCD corrections to $\mathbf{pp\to H^+W^-}$}
\label{QCDcorrections2}

The LO hadronic production cross section for the dominant gluon-fusion channel (in the heavy top-mass approximation) can be cast in the simple form
\be
\sigma_{\text{LO}} \; =\; \int_{\tau_0}^1 d\tau \int_\tau^1 \frac{dx}{x}  \;  g(x,\mu_F) \, g(\tau/x,\mu_F) \;  \hat{\sigma}_{\text{LO}}(\hat{s}=\tau s)\, ,
\label{crossPDFs2}
\ee
where $\hat{\sigma}_{\text{LO}}$ stands for the partonic cross section $\hat{\sigma}(gg\to H^+ W^-)$, given in Eq.~(\ref{sigmagg}), and $\tau_0=(M_{H^\pm}+M_W)^2/s$. At the NLO, the cross section can be written as \cite{DJ2,DJ1}
\be
\sigma_{\text{NLO}}\; =\; \sigma_{\text{LO}} + \Delta\sigma_{gg}^{\text{virt}}
+ \Delta\sigma_{gg} + \Delta\sigma_{q\bar{q}} + \Delta\sigma_{gq}\, ,
\ee
where:
\begin{align}
\Delta\sigma_{gg}^{\text{virt}} &\; =\;
 \frac{\alpha_s(\mu_R)}{\pi} \int_{\tau_0}^1 d\tau \int_\tau^1 \frac{dx}{x}  \;  g(x,\mu_F) \, g(\tau/x,\mu_F) \; \hat{\sigma}_{\text{LO}}(\tau s) \; \omega_{gg}^{\text{virt}} \, ,
\\[2ex]
\Delta\sigma_{gg} &\; =\;
 \frac{\alpha_s(\mu_R)}{\pi} \int_{\tau_0}^1 d\tau \int_\tau^1 \frac{dx}{x}  \;  g(x,\mu_F) \, g(\tau/x,\mu_F) \; \int_{\tau_0/\tau}^1 \frac{dz}{z} \; \hat{\sigma}_{\text{LO}}(\tau s z) \; \omega_{gg}(z) \, ,
\\[2ex]
\Delta\sigma_{gq} &\; =\; \frac{\alpha_s(\mu_R)}{\pi}  \int_{\tau_0}^1 d\tau \int_\tau^1 \frac{dx}{x}\; \sum_{q,\bar{q}}
  \; \Big[ \;
q(x,\mu_F) \, g(\tau/x,\mu_F)  +  g(x,\mu_F) \, q(\tau/x,\mu_F)
\notag \\
 & \qquad + \; \bar{q}(x,\mu_F) \, g(\tau/x,\mu_F)  +  g(x,\mu_F) \, \bar q(\tau/x,\mu_F) \;  \Big]  \; \int_{\tau_0/\tau}^1 \;  \frac{dz}{z} \; \hat{\sigma}_{\text{LO}} (\tau s z) \; \omega_{gq}(z) \, ,  
\end{align}

\begin{align}
\Delta\sigma_{q\bar{q}} &\; =\; \frac{\alpha_s(\mu_R)}{\pi}  \int_{\tau_0}^1 d\tau \int_\tau^1 \frac{dx}{x}\;\sum_{q,\bar{q}}
  \; \Big[ \;
q(x,\mu_F) \, \bar{q}(\tau/x,\mu_F)  +  \bar{q}(x,\mu_F) \, q(\tau/x,\mu_F) \;  \Big]
\notag \\
 & \qquad \times \; \int_{\tau_0/\tau}^1 \;  \frac{dz}{z} \; \hat{\sigma}_{\text{LO}} (\tau s z) \; \frac{32}{27}(1-z)^3 \, ,
\end{align}
with the functions $\omega_{gg}^{\text{virt}}$, $\omega_{gg}$ and $\omega_{gq}$ given by
\begin{align}
\omega_{gg}^{\text{virt}} &\; =\;  \pi^2 +  \frac{11}{2} + \frac{33-2N_f}{6}\,\log \Big(\frac{\mu_R^2}{\tau s} \Big)
\, , \notag \\[1.5ex]
\omega_{gg} &\; =\;  - z\, P_{gg}(z) \,  \log \Big(\frac{\mu_F^2}{\tau s}\Big) - \frac{11}{2}\, (1-z)^3
  + 12\, \Big(\frac{\log(1-z)}{1-z}\Big)_{+} - 12\, z\, (2-z+z^2)\,\log(1-z) \, , \notag \\[1.5ex]
\omega_{gq} &\; =\;  -\frac{z}{2}\, P_{gq}(z) \, \log \Big(\frac{\mu_F^2}{\tau s \, (1-z)^2}\Big) -1+2\, z-\frac{1}{3}\, z^2  \, ,
\end{align}
where $P_{gg}$ and $P_{gq}$ are the Altarelli-Parisi splitting functions
\begin{align}
P_{gg}(z) &\; =\; 6\, \Big[ \Big(\frac{1}{1-z}\Big)_{+} + \frac{1}{z} -2 + z\, (1-z) \Big] + \frac{33-2N_f}{6}\, \delta(1-z) \, ,
\notag \\[1.5ex]
P_{gq}(z) &\; =\; \frac{4}{3} \; \frac{1+(1-z)^2}{z}\, .
\end{align}


\end{appendix}

\section*{Acknowledgements}

This work has been supported in part by the Spanish
Government and ERDF funds from the EU Commission
[Grants FPA2011-23778 and CSD2007-00042
(Consolider Project CPAN)] and by Generalitat
Valenciana under Grant No. PROMETEOII/2013/007. The work of V.I. is supported by the Spanish Ministry MINECO through the FPI grant BES-2012-054676.

\end{document}